\documentclass[aps,twocolumn,prb,superscriptaddress,english,longbibliography]{revtex4-2}

\usepackage{amsmath}
\usepackage{amssymb}
\usepackage{amsthm}
\usepackage{makecell}
\usepackage{amsbsy}
\usepackage{amstext}
\usepackage{graphicx}
\usepackage{color}
\usepackage{float}

\usepackage{wasysym}

\usepackage{mathtools}
\usepackage{rotating}
\usepackage{mathtools}
\usepackage{tikz}
\makeatletter
\usepackage{amsfonts}
\usepackage{dcolumn}
\usepackage{bbold,bm}

\newcommand{\myblue}[1]{\textcolor[rgb]{0,.25,1}{#1}}
\newcommand{\myred}[1]{\textcolor[rgb]{1,0.078,0}{#1}}

\newcommand\vex[1]{\mathbf{#1}}
\newcommand\gvex[1]{\boldsymbol{#1}}

\def\id{\mathbb{1}} % need bbold,bm package

\makeatother
\graphicspath{{fig}}

\begin{document}

\title{Topological Floquet flat bands in irradiated alternating twist multilayer graphene}

\begin{abstract}
We study the appearance of topological Floquet flat bands in alternating-twist multilayer graphene, which has alternating relative twist angle $\pm\theta$ near the first magic angle. While the system hosts both flat bands and a steep Dirac cone in the static case, the circularly polarized laser beam can open a gap at the Moir\'{e} $K$ point and create Floquet flat bands carrying nonzero Chern numbers. Considering recent lattice-relaxation results, we find that the topological flat band is well-isolated for the effective interlayer tunneling in $n=3, 4, 5$ layers.  
Such dynamically produced topological flat bands are potentially observed in the experiment and thus provide a feasible way to realize the fractional Chern insulator.
\end{abstract}

\date{\today}

\author{Yingyi Huang}
\email{yyhuang@gdut.edu.cn}
\affiliation{School of Physics and Optoelectronic Engineering,
Guangdong University of Technology, Guangzhou 510006, China}

\maketitle

\section{Introduction}

Recent progress in twisted bilayer graphene (TBG) puts graphene back to the center of condensed matter physics because of the discovery of strong correlation and superconductivity in such systems~\cite{cao2018correlated,cao2018unconventional,yankowitz2019tuning}. These exotic phenomena appear to be related to the presence of flat bands, which is a result of the flattening of Dirac cones at certain special twist angles, called magic angles~\cite{bistritzer2011moire,dos2007graphene}.
 Around charge neutrality, interaction effects are enhanced by van Hove singularities coming from the nearly flat bands at these magic angles. On the other hand, the flat bands in TBG are generally topologically nontrivial, even in the absence of spin-orbit coupling~\cite{sharpe2019emergent,zhang2019twisted,bultinck2020ground,song2019all,song2021twisted}. A lot of studies were done to explore the topological feature in the mini Brillouin zone from the large Moir\'e superlattice~\cite{bultinck2020ground,lian2021twisted,zhang2019twisted,sharpe2019emergent,shen2021emergence}. These narrow enough topological flat bands are particularly relevant to 
various exotic fractional quantum Hall effects~\cite{abouelkomsan2020particle,repellin2020chern,wilhelm2021interplay,xie2021fractional}. However, the small magic angle and coupling ratio between intrasublattice and intersublattice hopping parameters make the realization of a fractional Chern insulator in TBG remains elusive.

Recently, people have turned their attention to multilayer graphene systems~\cite{mora2019flatbands,cea2019twists,zhu2020twisted,mao2023supermoire,lin2022energetic,ma2023doubled}. For example, alternating-twist multilayer graphene (ATMG) is a promising platform to realize phases seen in TBG~\cite{hao2021electric,cea2019twists,lake2021reentrant,qin2021plane,park2022robust,zhang2022promotion}, in which the nearest-neighboring layers are aligned and have alternating relative twists of $\pm \theta$. Remarkably, it can be mapped exactly to a sequence of decoupled TBG subsystems (plus single-layer graphene) for an odd (even) number of layers~\cite{ledwith2021tb,khalaf2019magic,popov2023magic}. Topological phases not only have been experimentally observed in other multilayer systems, including $ABC$ trilayer graphene on a hexagonal boron nitride~\cite{chen2019signatures,chen2020tunable,zhang2019nearly,liu2022magic,xie2022alternating} and twisted double-bilayer graphene~\cite{liu2020tunable,lee2019theory,cao2020tunable,shen2020correlated}, they also have been found in ATMG under an in-plane magnetic field and out-of-plane electric field~\cite{ledwith2021tb}. However, after decomposition, the intrasubsystem interaction of the electromagnetic fields vanishes in odd-layer ATMG or decays with increasing layer numbers in even-layer ATMG~\cite{ledwith2021tb}. To adjust the intrasubsystem coupling, it is necessary to utilize other experimental techniques. 

Optical engineering of the physical properties of a solid is a highly controllable method. In particular, strong circularly polarized light driving 
 opens the gap between bands by periodically changing the Hamiltonian and can be described by Floquet theory~\cite{Oka_2009,Gu_2011,Kitagawa_2011,Dora_2012,Iadecola_2013,Syzranov_2013,Ezawa_2013,
Kundu_2014,Grushin_2014,Kundu_2016,rudner2020band}. Its application can be dated back to the utilization of photons to open the gap in the Haldane model~\cite{sentef2015theory}.
Furthermore, the Floquet method has been generalized to TBG-related systems~\cite{li2020floquet,yang2023optical,hu2022floquet,vogl2020effective,vogl2020floquet,katz2020optically,topp2019topological,topp2021light,rodriguez2021low,rodriguez2020floquet,assi2021floquet,lu2021valley}. Interestingly, it has been found that Floquet engineering makes the low-energy flat band topological~\cite{li2020floquet,yang2023optical,hu2022floquet,vogl2020effective,vogl2020floquet,katz2020optically}. Thus, it is worthwhile to investigate ATMG under the driving of circularly polarized light.

\begin{figure*}[t]
   \centering
   \includegraphics[width=\linewidth]{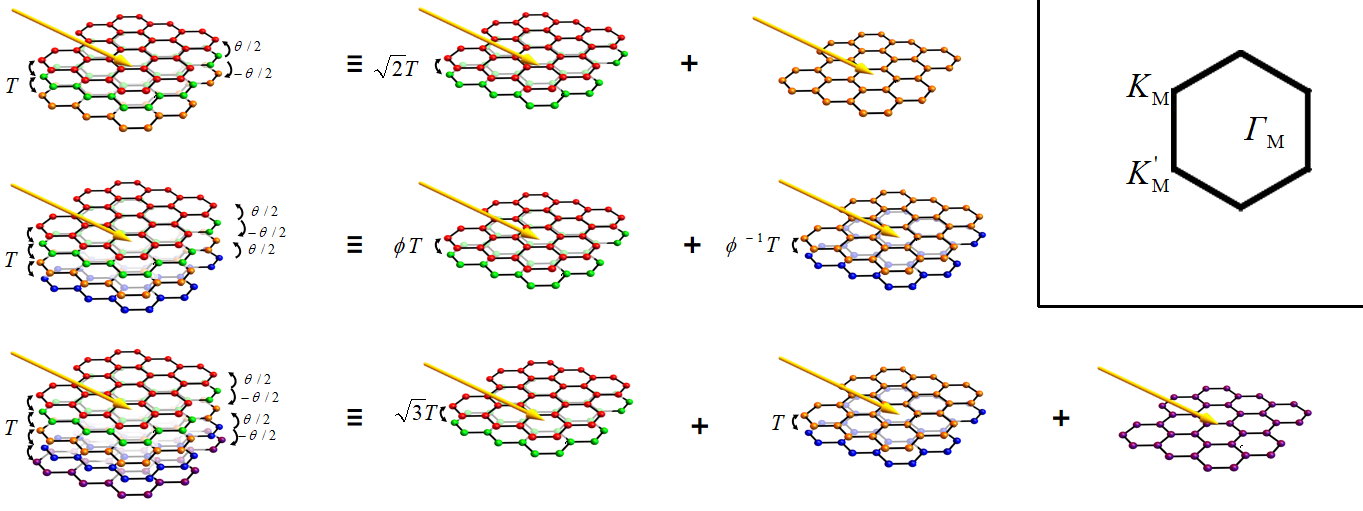} 
   \caption{Illustration of decomposition of ATMG irradiated by circularly polarized light $\vex A(t) = A (\cos \Omega t, \sin \Omega t)$ into TBG and graphene subsystems for $n=3, 4, 5$ layers. The light is marked by the yellow arrow. For $n=4$, the intersubsystem coupling ratio is $\phi=\frac{1+\sqrt{5}}{2}$ and its inverse. The terms generated by the light field act within subsystems. Inset: The Moir\'e Brillouin zone. }
   \label{fig:decompose}
\end{figure*}

In this paper, we study the effect of circularly polarized light on ATMG. 
Interestingly, we find that, under the high-frequency limit, irradiated ATMG can be mapped to irradiated bilayer subsystems (plus a irradiated single layer), as illustrated in Fig.~\ref{fig:decompose}. This mapping is similar to those with an electric field or a magnetic field~\cite{ledwith2021tb,khalaf2019magic,popov2023magic}, but there is no decay factor rescaling the light field. In particular, despite the existence of a Dirac cone at the $K_M$ point, the light field induces gap opening and isolates the central Floquet band (see Fig.~\ref{fig:gap}), and thus changes the topology of the flat band. 
The model is given in
Sec.~II and the mapping is presented in Sec.~III. In Sec.~IV, we calculate the Floquet band spectrum in trilayer graphene and show the existence of the Floquet topological flat band. The numerical results for $n=4, 5$ layers are presented in Sec.~V. Since the coupling ratio $u$ is layer dependent and sensitive to lattice relaxation in ATMG~\cite{ledwith2020fractional,ledwith2021tb}, we check the existence of Floquet topological flat band at the first magic angle and its corresponding effective $u$ in different layers. The advantages of the Floquet method on ATMG are addressed in the Conclusion in Sec.~VI.
%%%%%% Fig. 1 Setup %%%%%%%

\section{Model}
Now we consider a general system of $n$-layer ATMG irradiated by a beam of circularly polarized light. The relative twist between two neighboring layers has the same magnitude but alters in sign ($\pm\theta$). Here, the relative displacement is neglected since the zero shift configuration is energetically favorable~\cite{ledwith2021tb}. The electromagnetic vector potential of the light is given by $\vex A(t) = A (\cos \Omega t, \sin \Omega t)$ with $t$ the time, $\Omega$ the frequency, and $A$ the field strength. The multiplication of $\Omega$ and $A$ is the electric field amplitude, $E=\Omega A$.

The low-energy physics of the system for one of the valleys can be modeled by
%%%%%
\begin{equation}\label{eq:FTBG}
H(t) = \begin{bmatrix}
h_{\theta/2}(t) & \mathfrak{T} & 0 & \cdots \\
\mathfrak{T}^\dagger & h_{-\theta/2}(t) & \mathfrak{T}^\dagger& \cdots\\
0 & \mathfrak{T} & h_{\theta/2}(t)& \cdots \\
\cdots & \cdots & \cdots & \cdots
\end{bmatrix},
\end{equation}
%%%%%
where $h_{\theta/2}(t) = v_F[-i\hbar \gvex\nabla - e \vex A(t)]\cdot\gvex\sigma_{\theta/2}$, with rotated Pauli matrices $\gvex\sigma_{\theta/2} \equiv e^{-i\theta\sigma_z/4} (\sigma_x,\sigma_y) e^{i\theta\sigma_z/4}$ and Fermi velocity $v_F$, is the low-energy Dirac Hamiltonian of a valley of a single graphene sheet twisted by angle $\theta/2$. The time-dependent electromagnetic potential $\vex A(t)$ is introduced into the Hamiltonian by way of minimal substitution. 
The interlayer tunneling matrix $\mathfrak{T} = \sum_{n=1}^3 \mathfrak{T}_n e^{-i k_\theta \vex q_n \cdot \vex r}$, with
%%%%%
\begin{equation}
\frak{T}_n = w_{AA} \sigma_0 + w_{AB} \vex q_n \cdot \gvex\sigma_{\pi/2},
\end{equation}
%%%%%
where the unit vectors $\vex q_1 = (0,-1)$, $\vex q_{2,3} =  (\pm\sqrt3/2,1/2)$ encode the tunneling $w_{AA}$ and $w_{AB}$ between the $AA$- and $AB$-stacked regions of the TBG. $k_\theta = 8\pi \sin(\theta/2)/3 a$ is the wave vector of the Moir\'e pattern and $a$ is the Bravais lattice spacing of graphene.

 In the Floquet theory, the periodically changed vector potential modifies the static Hamiltonian to $H_{\vex k}(t) = H_{\vex k}(t+T)$. One focuses on quasienergies $\epsilon_{\vex k s}$, and periodically changed Floquet modes $|\phi_{\vex k s}(t)\rangle = |\phi_{\vex k s}(t+T)\rangle$, where $T$ is the period and ${\bf k}$ is the crystal momentum, by solving the Floquet-Schr\"odinger equation $[H_{\vex k}(t) - i\hbar\partial_t]|\phi_{\vex k s}(t)\rangle = \epsilon_{\vex k s} |\phi_{\vex k s}(t)\rangle$. The quasienergies fall into a so-called Floquet zone with size $\hbar\Omega$ similar to the concept of the Brillouin zone but in the time dimension. The relation between Floquet modes $|\phi_{\vex k s}(t)\rangle$ and the wave function governed by the time-dependent Schr\"odinger equation $|\phi_{\vex k s}(t)\rangle$ is $|\phi_{\vex k s}(t)\rangle\equiv e^{i \epsilon_{\vex k s} t /\hbar}|\psi_{\vex k s}(t)\rangle$.
 
The simulation parameters are chosen from the experimentally known electronic structure~\cite{kuzmenko2009determination} as follows: $a= 2.4$~\AA, $\hbar v_F / a = 2.425$~eV, and $w_{AB} = 112$~meV . Correspondingly, $\alpha \equiv w_{AB}/{\hbar v_F k_\theta} = 1.1\times 10^{-2} / 2\sin(\theta/2)$ is a function of the twist angle $\theta$.  Throughout this paper, the laser parameters are chosen to be $A=0.08a^{-1}$ and $\Omega=6eV/\hbar$, which are experimentally attainable.  Under this circumstance, the relevant energy scales of the low-lying bands are much lower than the Floquet sidebands, and thus the admixture with the Floquet sideband can be neglected in our following discussions.

The interlayer tunneling around the magic angle is affected by atomic relaxation. Especially, this relaxation effect becomes stronger with the number of layers being increased~\cite{ledwith2021tb,leconte2022electronic}. The relaxation of atoms in multilayer geometry will decrease the interlayer distance at $AB$ stacking and increase the one at $AA$ stacking. This reduces the $AA$ tunneling $w_0$ and increases the $AB$ tunneling $w_1$, and thus changing their ratio $u=w_0/w_1$. In trilayer graphene ($n=3$), the first-principles calculation gives the value of the first magic angle $\theta=1.49^\circ$ and the effective $u=0.585$. For $n=4, 5$, the distinction between tunneling at inner and outer interfaces changes the value of the magic angle. For $n=4$, the value of the first magic angle is $\theta=1.68^\circ$ and the effective $u=0.614$. For $n=5$, the second magic angle is within reach. The value of the first and second magic angles are $\theta=1.79^\circ$ and $\theta=1.14^\circ$, and the corresponding effective $u$ are $0.627$ and $0.45$~\cite{ledwith2021tb}.

\section{\label{sec:decompose}The reduction of effective Hamiltonian in the high-frequency limit}
In the high-frequency limit, light does not directly excite electrons and instead effectively modifies the electron band structures. Its influence can be represented by an effective static Hamiltonian $\Delta H_{F}=i\hbar/T\log U$, where $U=\mathcal{T}\exp[-i/\hbar\int^T_0 H(t)dt]$ is the time-evolution operator, with $\mathcal{T}$ being the time-ordering operator. 
For $A^2\ll 1$, we consider the  $\delta H_{F}= [H^{(-1)},H^{(1)}]/\hbar\Omega$. Here, $H^{(n)} = \int_0^1 e^{-2\pi in \tau}H(2\pi\tau/\Omega) d\tau$
for $n=0,1,2,\cdots$ are the Fourier components of the periodic Hamiltonian.

Now we have the effective Hamiltonian $H_F=H^{(0)}+\delta H_F$, where $H^{(0)}$ the static part with $A$ being set to zero and the modification part being
\begin{equation}
\delta H_F=B(\id\otimes\sigma_z).
\label{eq:sigzterm}
\end{equation}
Here, $B=\frac{(ev_F A)^2}{\hbar \Omega}$ and $\id$ is an $n$-dimensional identity matrices for an $n$-layer system. $\sigma_z$ is a Pauli matrix acting on the individual layer's sublattice degree of freedom, and thus breaks time-reversal symmetry.

Following a similar procedure in static systems~\cite{khalaf2019magic}, we continue to decompose the effective Hamiltonian $H_{F}$ for $n=3,4,5$ layers irradiated ATMG using a unitary transformation. The transformation $V$ is a $2n$-dimensional transformation matrix, which is given in Refs.~\onlinecite{khalaf2019magic,ledwith2020fractional}. Using the unitary transformation on the static part and the modification part of $H_{F}$ separately, we have 
$V^T(H^{(0)}+\delta H_{F})V=V^TH^{(0)}V+V^T\delta H_{F}V $. 

After the transformation, the static Hamiltonian generates a TBG subsystem $\mathcal{H}_1$. For different layers of ATMG, the remaining subsystems $\mathcal{H}_2, \mathcal{H}_3, \cdots$ that arise from the static Hamiltonian can be graphene, non-TBG, or their combinations. For the modification part, since $V$ is independent of the sublattices, the transformation leaves $\delta H_{F}$ unchanged. That is, $V^T\delta H_{F}V=\delta H_{F}$. From the fact that the modification parts can be written as a direct sum of $\sigma_z$ terms, $\delta H_{F}=B(\sigma_z\oplus\sigma_z\oplus\cdots)$, we have    
\begin{equation}
    H^{dec}=(\mathcal{H}_1+B\sigma_0\otimes\sigma_z)\oplus(\mathcal{H}_2+B\sigma_0\otimes\sigma_z)\oplus\cdots.
    \label{eq:Hdec}
\end{equation}
For TBG, the corresponding term becomes $\mathcal{H}_i+B\sigma_0\otimes\sigma_z$.
For single-layer graphene, the corresponding term becomes $\mathcal{H}_i+B\sigma_z$.
From Eq.~\eqref{eq:Hdec}, the effect of a beam of circularly polarized light on the ATMG can be presented by multiple independent beams of circularly polarized light on each subsystem, as shown in Fig.~\ref{fig:decompose}.

%%%%%%%%%%%%%%%%%%%%%%%%%fig 2
\begin{figure}
    \centering   
\includegraphics[width=1.0\linewidth]{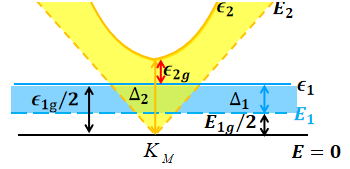}
    \caption{\label{fig:gap}The schematics of low-energy band structure of $n=3$ ATMG near $K_M$ point at a magic angle. The static (Floquet) bands $E's$ ($\epsilon's $) are represented by dashed (solid) lines. The first (second) lowest energy level $E_{1}$ and $\epsilon_{1}$ ($E_{2}$ and $\epsilon_{2}$) in blue (yellow) are from the TBG (graphene) subsystem. The gap opening of the first (second) lowest energy level at $K_M$ point is marked by the blue (yellow) region $\Delta_{1(2)}=\epsilon_{1(2)}-E_{1(2)}$. The distance from the lowest static (Floquet) bands to zero energy is denoted by $E_{1g}/2$ ($\epsilon_{1g}/2$). The gap between the two static (Floquet) bands is denoted by $E_{2g}$ ($\epsilon_{2g}$).Note that particle-hole asymmetry is not considered here.}   
\end{figure}
One of the remarkable consequences is that the $\delta H_{F}$ modifies the low-energy electron band structure. To illustrate the effect of light, we consider a three-layer ATMG as shown in Fig.~\ref{fig:gap}. We focus on the two lowest positive energy bands. For static bands ($E's$), while the lowest one ($E_1$) is a flat band in the TBG subsystem, the second one from the other subsystem ($E_2$) hosts a Dirac cone at the $K_M$ point. The effect of light on the two subsystems can be shown by the shifts of the two lowest Floquet energy levels ($\epsilon's$). 
Actually, as evident in Fig.~\ref{fig:gap}, the gap openings in the low-energy subsystems at the $K_M$ point isolate the central Floquet band $\epsilon_1$. 
The band gap between the positive and negative central Floquet bands $\epsilon_{1g}$ is nonzero and is twice the distance between $\epsilon_1$ and zero energy $\epsilon_{1g}/2=E_{1g}/2+\Delta_1$, with $\Delta_1$ being the gap opening of the lower energy level in TBG subsystem.
For the band gap above the positive central Floquet band $\epsilon_{2g}$, we should discuss $u=0$ and $u>0$ cases separately. In the chiral limit ($u=0$), the static bands $E_1$ and $E_{2}$ touch each other at zero energy, i.e., $E_{1g}=0$. 
The Floquet band gap $\epsilon_{2g}=\Delta_2-\Delta_1$. The shift of the Dirac cone under $B\sigma_z$ term is $\Delta_2=B$. $\Delta_1$ can be calculated from the eigenenergies of $\delta H_F$ as $B$. Thus, the $\epsilon_{2g}$ gap keeps closing at $K_M$ points for $u=0$.
In contrast, for $u>0$, $\epsilon_{2g}=\Delta_2-\Delta_1-E_{1g}/2$. $\Delta_2$ is independent of $u$ and always equal to $B$.  In TBG system, $\epsilon_{1g}=2\Delta_1+E_{1g}$ is always smaller than the value at zero $u$~\cite{li2020floquet}, which means $\epsilon_{1g}<B$. Therefore, for finite $u$, $\Delta_1+E_{1g}/2$ is smaller than $\Delta_2$ and leads to the opening of gap $\epsilon_{2g}$.

From the above analysis, we can see that the polarized light on the ATMG opens the gaps $\epsilon_{1g}$ and $\epsilon_{2g}$ at the $K_M$ point and isolate the flat band with $u>0$. This can be generalized to ATMG with $n>3$. 

 Actually, the phase difference $\phi_0$ between the $x$ and $y$ components of the circularly polarized light can be different, which is the elliptically polarized light case $\mathbf{A}(t)=(A_x cos\Omega t, A_y cos(\Omega t+\phi_0))$. In this case, the above analysis can also be true with the factor in Eq.~\eqref{eq:sigzterm} being $B=\frac{(ev_F)^2A_xA_y}{\hbar \Omega}\sin\phi_0$, except for the linearly polarized case in which $\phi_0=0$ leads to the vanishing of the $\delta H_F$ term.

\section{The trilayer case ($n=3$)}
To check the isolation of the central Floquet band, we now examine the $n=3$ ATMG Floquet spectrum numerically.

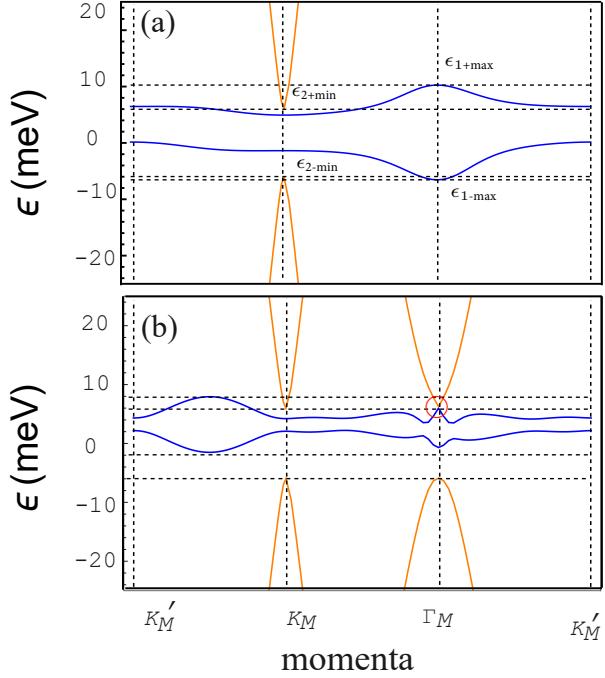
\begin{figure}
    \centering   
\includegraphics[width=1.0\linewidth]{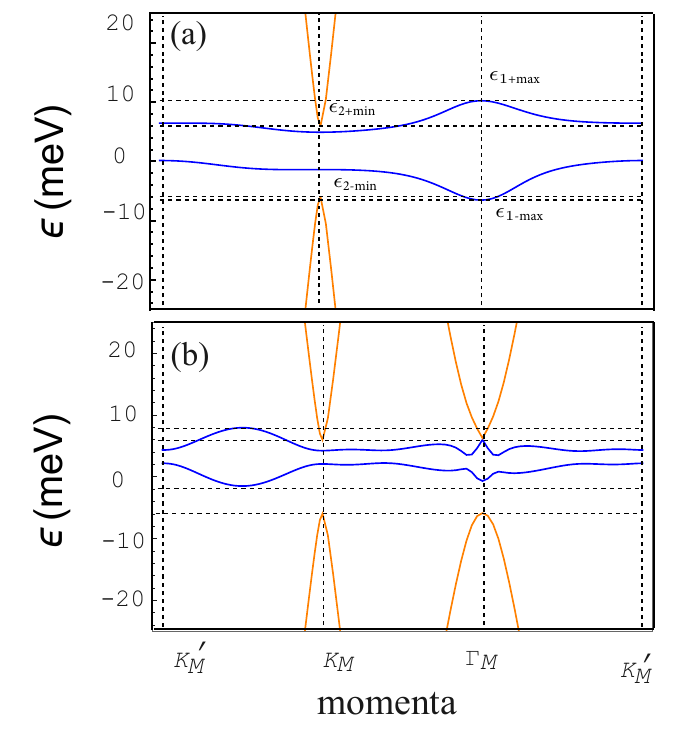}
  \begin{center}
 \begin{picture}(1,1) 
  \put(51,123){\myred{\circle{8}}}
\end{picture}
\end{center}
\caption{(a)The band structure of irradiated trilayer ATMG at the first magic angle $\alpha=0.425$ and the corresponding effective interlayer tunneling $u=0.585$, which is pointed out by a red star in Fig.~\ref{fig:3layer}(c). The central bands are nontrivial and have Chern number $\pm 4$. (b) The band structure of irradiated trilayer ATMG at a gap closing point $(\alpha, u)=(0.5, 0.8024)$, which is pointed out by a black arrow in Fig.~\ref{fig:3layer}(d). The red circle marks the $\epsilon_{2+g}$ gap closing, indicating the central band $\epsilon_{1+}$ trivial. The laser frequency and electric field are set at $\hbar\Omega=6eV$ and $E=2\times 10^4 kV/cm$. 
}
    \label{fig:spectrum}
\end{figure}

The Floquet spectrum can be calculated numerically by the Hamiltonian Eq.~\eqref{eq:FTBG} via the Floquet-Schrödinger approach. The Floquet spectrum shown in Fig.~\ref{fig:spectrum}(a) is at the first magic angle and the corresponding tunneling ratio chosen from the lattice relaxation result. The spectrum exhibits apparent electron-hole asymmetry due to the relaxation~\cite{song2021twisted}. Similar to the TBG case, the hole side gets much wider than the electron side. The positive (negative) central Floquet energy bands $\epsilon_{1+(-)}$ corresponds to the lower band of the TBG subsystem. And the Floquet band next to the central one $\epsilon_{2+(-)}$ corresponds to the graphene subsystem and hosts a steep Dirac cone at $K_M$ point, which is absent in TBG system. 

While the gap opening at the $K_M$ point for finite $u$ is proved in the last section, the indirect gap between the minimum value of $\epsilon_{2+}$ and the maximal value of $\epsilon_{1+}$ is not always opened. Figure \ref{fig:spectrum}(a) shows that the minimal of $\epsilon_{2+}$ at $\Gamma_M$ point is lower than the maximal value of  $\epsilon_{1+}$ at $K_M$ point, giving a negative indirect gap. Thus, we should use the direct gap instead of the indirect one to characterize the band feature. 

A numerical calculation of Chern numbers at $(u, \alpha)=(0.585, 0.425)$ [Fig.~\ref{fig:spectrum} (a)] gives a nontrivial topological number $\pm 4$. This can be understood from the fact that time-reversal breaking induces valley and spin degeneracies in the irradiation ATMG system, similar to the TBG case~\cite{li2020floquet}.
In Fig.~\ref{fig:spectrum}(b), for $(u, \alpha)=(0.5, 0.8024)$, however, a gap closing takes place at the $\Gamma_M$ points and gives a trivial Chern number. Thus, it is necessary to investigate the band features under different effective interlayer tunneling $u$ and twist angle $\alpha$.

\begin{figure}
   % \centering   
\includegraphics[width=1.0\linewidth]{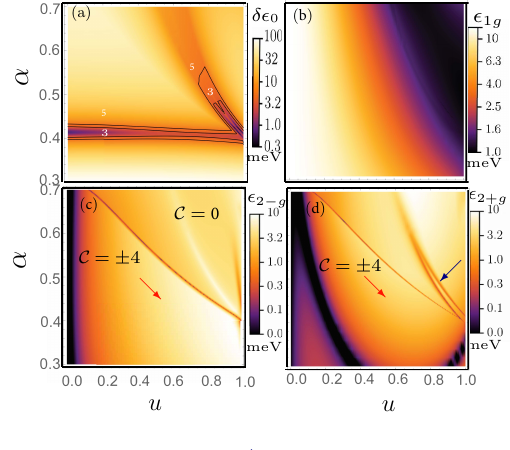}
\begin{center}
 \begin{picture}(1,1) 
 \put(63,69){\Large\myred{$\star$}}
  \put(-46,69){\Large\myred{$\star$}}
 \end{picture}
 \end{center}
 
\caption{(a) The bandwidth for the central band $\epsilon_0$, (b) the Floquet band gaps between central bands $\epsilon_{1g}$, (c) the Floquet band gaps between the central negative band and the next band $\epsilon_{2-g}$, and (d) the Floquet band gaps between the central positive band and the next band $\epsilon_{2+g}$ as a function of twist angle and the ratio of interlayer tunneling $u$. In (c) and (d), the red star denotes the first magic angle and its corresponding $u$ from lattice relaxation. The laser parameters are the same as in Fig.~\ref{fig:spectrum}.}
\label{fig:3layer}
\end{figure}
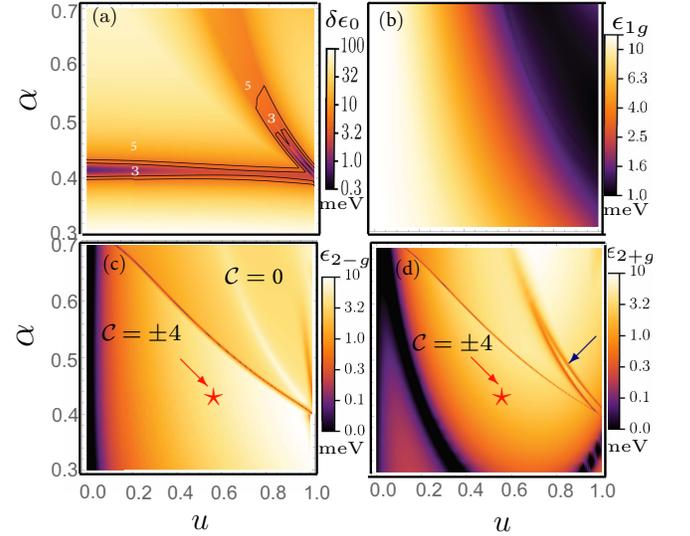

Now we show the phase diagrams for the bandwidth and band gaps of the central Floquet band.
The bandwidth of the two central bands is shown in Fig.~\ref{fig:3layer}(a), sharing many similarities to the irradiated TBG. In particular, the irradiated ATMG has advantages over static cases. First, the flat bands exist over a wider range of twist angles. This is true in both regions: the one around the magic angle ($\alpha=0.425$) in the whole range of $u$ and the one at $\alpha>0.425$ and large $u$. Second, the bandwidths of the lowest energy bands in the irradiated case are smaller than those in the static case. This is obvious near the chiral limit ($u<0.2$) and small twist angle ($\alpha>0.5$). The most important characteristic of $n=3$ ATMG is that the flat band regions are at larger twist angles than irradiated TBG. This can be understood from the fact that the magic angle of three-layer ATMG is $\sqrt{2}\theta_{TBG}$, which is larger than the one of TBG.

Before searching the topological region of the flat band, we should find the band isolation regions by calculating the band gaps.
The gap $\epsilon_{1g}$ can be seen in Fig.~\ref{fig:3layer}(b). It becomes smaller with increasing $u$ but remains non-vanishing in the entire regions. This is consistent with the TBG case since the $\epsilon_{1+(-)}$ bands belong to the TBG subsystems. 

The situation for the $\epsilon_{2g}$ gap is more complicated. Due to the electron-hole asymmetry, the $\epsilon_{2+g}$ and $\epsilon_{2-g}$ gaps for the electron and hole parts are not the same. Figure~\ref{fig:3layer}(c) exhibits fewer dark regions which correspond to fewer gap closings. This can be understood from the fact that the hole part of the spectrum is wider than the electron part.

For the $\epsilon_{2-g}$ gap, we found two gap closings. The first gap closing at $K_M$ point is around the chiral limit $u=0$, as discussed in Sec.~\ref{sec:decompose}. As the interlayer tunneling ratio increases, another gap closing appears at the $\Gamma_M$ point for large $u$, as given in Fig.~\ref{fig:spectrum}(b). The region between the two gap closings is topological. Remarkably, the flat band at the first magic angle and its corresponding effective interlayer ratio from lattice relaxation calculation falls into this topological region (marked by a red star).

For the positive energy, the phase diagram of $\epsilon_{2+g}$ [Fig.~\ref{fig:3layer}(d)] exhibits two additional gap closings at the $\Gamma_M$ point for smaller twist angles closing to $u=1$. They are very close to the gap closings in the TBG case (see Fig.~3(b) in Ref.~\cite{li2020floquet}) with a $\sqrt{2}$ amplification factor on the $y$ axis. 

Therefore, although the gap closings at the $\Gamma_M$ point make the phase diagram for $n=3$ ATMG more complicated, they have no influence on the existence of the topological isolated flat bands at the first magic angle, whose twist angle is larger than the one of TBG. 

%%%%%%%%%%%%%%%%%%%%%%%%%%%%%%%fig 3
%%%%%%%%%%%%%%%%%%%%%%%%fig 4

\section{$n=4,5$}
For an ATMG system with more layers, more subsystems appear after decomposition. We should look at their influences on the central Floquet bandwidth and bandgaps.

\begin{figure}
    \centering
    \includegraphics[width=1.0\linewidth]{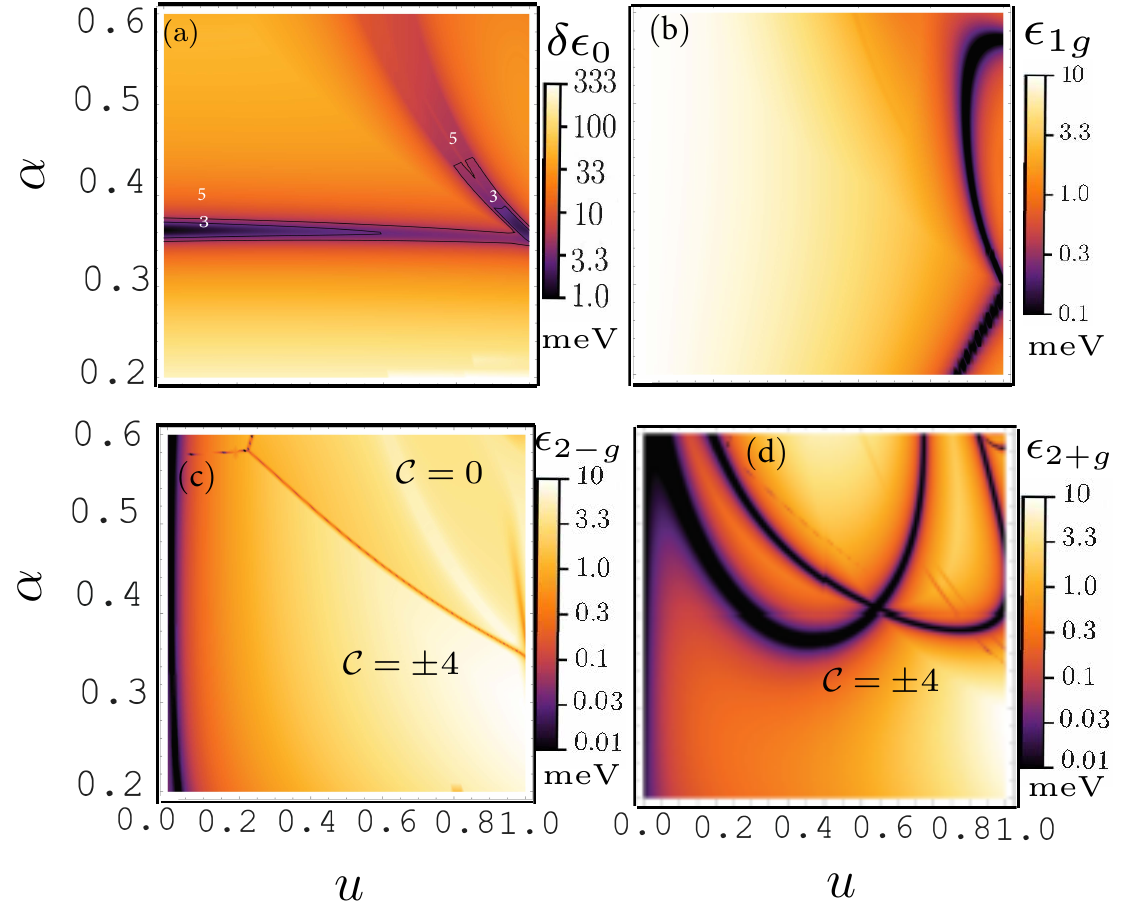}
  \begin{center}
 \begin{picture}(1,1) 
 \put(68,75){\Large\myred{$\star$}}

  \put(-43,77){\Large\myred{$\star$}}
   
 \end{picture}
 \end{center}
    \caption{The bandwidth and band gap for $n=4$ layer. The labels and laser parameters are the same as in Fig.~\ref{fig:3layer}.}
    \label{fig:4layer}
\end{figure}
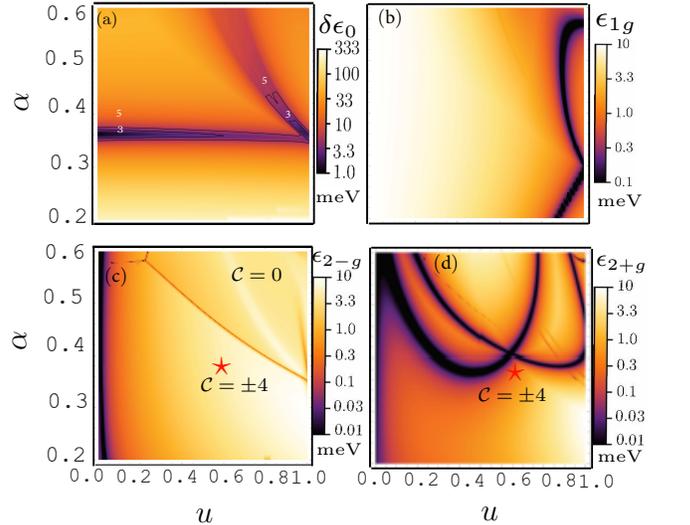

\begin{figure}
    \centering
    \includegraphics[width=1.0\linewidth]{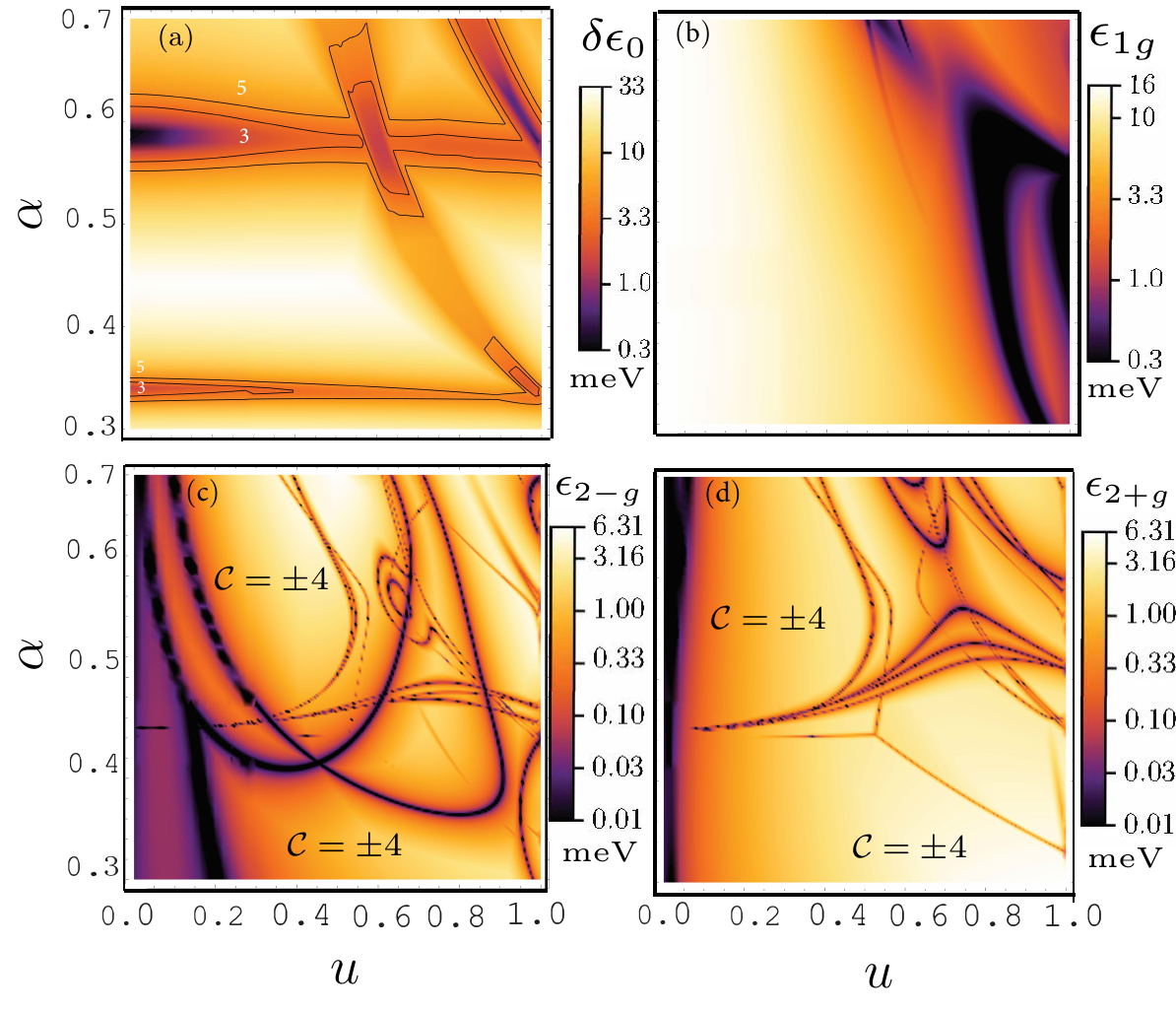}
  \begin{center}
 \begin{picture}(1,1) 
 \put(51,98){\Large\myblue{$\star$}}
 \put(65,55){\Large\myred{$\star$}}
 \put(-60,98){\Large\myblue{$\star$}}
  \put(-44,55){\Large\myred{$\star$}}

 \end{picture}
 \end{center}
    \caption{The bandwidth and band gap for $n=5$ layer. The labels and laser parameters are the same as in Fig.~\ref{fig:3layer}. In (c) and (d), the red and blue stars denote the first and the second magic angles and their corresponding $u$ from lattice relaxation. }
    \label{fig:5layer}
\end{figure}
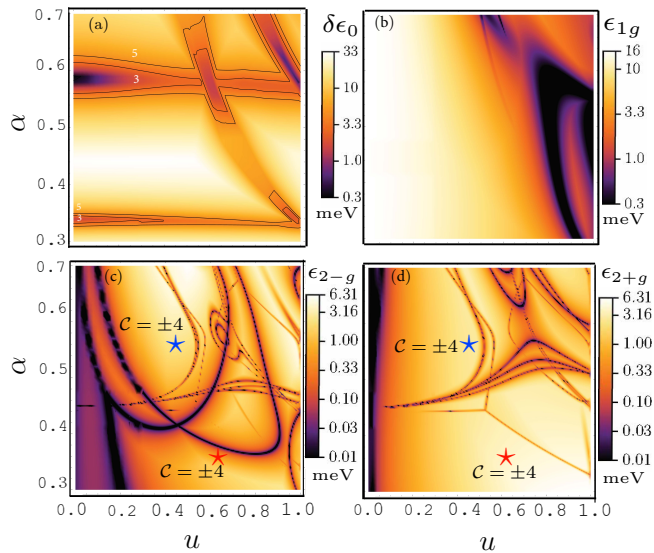

According to the discussion in Sec.~\ref{sec:decompose}, we know that the laser field opens gaps in the TBG subsystem for ATMG with different layer numbers. Most importantly, for ATMG, except the interlayer tunneling, both the laser parameter $B=\frac{(ev_FA)^2}{\hbar\Omega}$ and intralayer tunneling in the TBG subsystem do not depend on the layer number. We expect that the bandwidth $\delta\epsilon_0$ and the band gap $\epsilon_{1g}$ phase diagram in Figs.~\ref{fig:4layer}(a), \ref{fig:4layer}(b), \ref{fig:5layer}(a), \ref{fig:5layer}(b) are similar to the bilayer and the trilayer cases.

Actually, the increment of subsystems changes the $\epsilon_{2g}$ gap. The $\epsilon_{2-g}$ gap for $n=4$ and $n=5$ are shown in Figs.~\ref{fig:4layer}(c) and \ref{fig:5layer}(c). As a result of particle-hole asymmetry, the $\epsilon_{2+g}$ gap is more complicated and is shown in Figs.~\ref{fig:4layer}(d) and \ref{fig:5layer}(d).
For $n=4$, the ATMG decomposes into a TBG subsystem and a nonmagic TBG subsystem. The nonmagic TBG subsystem, which is away from the magic angles, hosts Dirac cones at $K_M$ and $K_M'$ points. When these Dirac cones touch the lowest energy level $\epsilon_1$, they close the band gap $\epsilon_{2-g}$. This is similar to the effect of the Dirac cone of the graphene subsystem in the trilayer system. In Fig.~\ref{fig:4layer}(d), we can see that the $\epsilon_{2+g}$ gap closing changes the phase diagram. However, the $\epsilon_{2+g}$ gap at the first magic angle is opened, and thus the central band is topologically nontrivial.

For the $n=5$ case, there are a non-TBG subsystem and a graphene subsystem besides the TBG subsystem. The appearance of Dirac cones in either the nonmagic TBG or the graphene subsystem closes the $\epsilon_{2g}$ gap. As a result, a more complicated phase diagram is found in Figs.~\ref{fig:5layer}(c) and \ref{fig:5layer}(d). The bands at the two first magic angles are in the gapped and topological phases.

%%%%%%%%%%%%%%%%%%%%%%%%fig 5
%%%%%%%%%%%%%%%%%%%%%%%fig6

\section{Discussion and Conclusion}

In this paper, we have investigated irradiated ATMG at charge neutral. By mapping the multilayer system into a direct sum of bilayer systems (plus a single-layer system), we have shown that the laser field opens gaps between the central Floquet flat bands and between the ones next to them for $u>0$. Numerical results of $n=3, 4, 5$ further show that the gap opening makes the Floquet flat bands topological in certain twist angle and tunneling ratio regions despite the complicated gap closing features at $K_M$ and $\Gamma_M$ points induced by the coexisting Dirac cones. These findings extend the Floquet study on TBG and other multilayer graphene systems, confirming the existence of the topological Floquet flat band in ATMG.

Floquet engineering ATMG is a promising platform to realize the Floquet fractional Chern insulators, which can be seen in two ways. 
First, compared to other multilayer systems studied in the literature, the flat band in ATMG is more stable. Distinguishing from $ABC$ trilayer graphene stacked on hexagonal boron nitride or twisted double bilayer graphene, does not exhibit magic angles or flat bands when realistic effects, e.g., trigonal warping terms, are included.
Most notably, we study the ATMG by decomposing it into several subsystems and find that the coupling of the laser field to the system is intra-subsystem and does not decay with the increasing layer number, which is more controllable to open gaps than other techniques. 

\begin{acknowledgments}
This work is supported by the National Natural Science Foundation of China (Grant No. 12104099).
\end{acknowledgments}

\bibliography{multilayerBG.bib}

\begin{thebibliography}{66}%
\makeatletter
\providecommand \@ifxundefined [1]{%
 \@ifx{#1\undefined}
}%
\providecommand \@ifnum [1]{%
 \ifnum #1\expandafter \@firstoftwo
 \else \expandafter \@secondoftwo
 \fi
}%
\providecommand \@ifx [1]{%
 \ifx #1\expandafter \@firstoftwo
 \else \expandafter \@secondoftwo
 \fi
}%
\providecommand \natexlab [1]{#1}%
\providecommand \enquote  [1]{``#1''}%
\providecommand \bibnamefont  [1]{#1}%
\providecommand \bibfnamefont [1]{#1}%
\providecommand \citenamefont [1]{#1}%
\providecommand \href@noop [0]{\@secondoftwo}%
\providecommand \href [0]{\begingroup \@sanitize@url \@href}%
\providecommand \@href[1]{\@@startlink{#1}\@@href}%
\providecommand \@@href[1]{\endgroup#1\@@endlink}%
\providecommand \@sanitize@url [0]{\catcode `\\12\catcode `\$12\catcode
  `\&12\catcode `\#12\catcode `\^12\catcode `\_12\catcode `\%12\relax}%
\providecommand \@@startlink[1]{}%
\providecommand \@@endlink[0]{}%
\providecommand \url  [0]{\begingroup\@sanitize@url \@url }%
\providecommand \@url [1]{\endgroup\@href {#1}{\urlprefix }}%
\providecommand \urlprefix  [0]{URL }%
\providecommand \Eprint [0]{\href }%
\providecommand \doibase [0]{https://doi.org/}%
\providecommand \selectlanguage [0]{\@gobble}%
\providecommand \bibinfo  [0]{\@secondoftwo}%
\providecommand \bibfield  [0]{\@secondoftwo}%
\providecommand \translation [1]{[#1]}%
\providecommand \BibitemOpen [0]{}%
\providecommand \bibitemStop [0]{}%
\providecommand \bibitemNoStop [0]{.\EOS\space}%
\providecommand \EOS [0]{\spacefactor3000\relax}%
\providecommand \BibitemShut  [1]{\csname bibitem#1\endcsname}%
\let\auto@bib@innerbib\@empty
%</preamble>
\bibitem [{\citenamefont {Cao}\ \emph {et~al.}(2018{\natexlab{a}})\citenamefont
  {Cao}, \citenamefont {Fatemi}, \citenamefont {Demir}, \citenamefont {Fang},
  \citenamefont {Tomarken}, \citenamefont {Luo}, \citenamefont
  {Sanchez-Yamagishi}, \citenamefont {Watanabe}, \citenamefont {Taniguchi},
  \citenamefont {Kaxiras} \emph {et~al.}}]{cao2018correlated}%
  \BibitemOpen
  \bibfield  {author} {\bibinfo {author} {\bibfnamefont {Y.}~\bibnamefont
  {Cao}}, \bibinfo {author} {\bibfnamefont {V.}~\bibnamefont {Fatemi}},
  \bibinfo {author} {\bibfnamefont {A.}~\bibnamefont {Demir}}, \bibinfo
  {author} {\bibfnamefont {S.}~\bibnamefont {Fang}}, \bibinfo {author}
  {\bibfnamefont {S.~L.}\ \bibnamefont {Tomarken}}, \bibinfo {author}
  {\bibfnamefont {J.~Y.}\ \bibnamefont {Luo}}, \bibinfo {author} {\bibfnamefont
  {J.~D.}\ \bibnamefont {Sanchez-Yamagishi}}, \bibinfo {author} {\bibfnamefont
  {K.}~\bibnamefont {Watanabe}}, \bibinfo {author} {\bibfnamefont
  {T.}~\bibnamefont {Taniguchi}}, \bibinfo {author} {\bibfnamefont
  {E.}~\bibnamefont {Kaxiras}}, \emph {et~al.},\ }\bibfield  {title} {\bibinfo
  {title} {Correlated insulator behaviour at half-filling in magic-angle
  graphene superlattices},\ }\href@noop {} {\bibfield  {journal} {\bibinfo
  {journal} {Nature}\ }\textbf {\bibinfo {volume} {556}},\ \bibinfo {pages}
  {80} (\bibinfo {year} {2018}{\natexlab{a}})}\BibitemShut {NoStop}%
\bibitem [{\citenamefont {Cao}\ \emph {et~al.}(2018{\natexlab{b}})\citenamefont
  {Cao}, \citenamefont {Fatemi}, \citenamefont {Fang}, \citenamefont
  {Watanabe}, \citenamefont {Taniguchi}, \citenamefont {Kaxiras},\ and\
  \citenamefont {Jarillo-Herrero}}]{cao2018unconventional}%
  \BibitemOpen
  \bibfield  {author} {\bibinfo {author} {\bibfnamefont {Y.}~\bibnamefont
  {Cao}}, \bibinfo {author} {\bibfnamefont {V.}~\bibnamefont {Fatemi}},
  \bibinfo {author} {\bibfnamefont {S.}~\bibnamefont {Fang}}, \bibinfo {author}
  {\bibfnamefont {K.}~\bibnamefont {Watanabe}}, \bibinfo {author}
  {\bibfnamefont {T.}~\bibnamefont {Taniguchi}}, \bibinfo {author}
  {\bibfnamefont {E.}~\bibnamefont {Kaxiras}},\ and\ \bibinfo {author}
  {\bibfnamefont {P.}~\bibnamefont {Jarillo-Herrero}},\ }\bibfield  {title}
  {\bibinfo {title} {Unconventional superconductivity in magic-angle graphene
  superlattices},\ }\href@noop {} {\bibfield  {journal} {\bibinfo  {journal}
  {Nature}\ }\textbf {\bibinfo {volume} {556}},\ \bibinfo {pages} {43}
  (\bibinfo {year} {2018}{\natexlab{b}})}\BibitemShut {NoStop}%
\bibitem [{\citenamefont {Yankowitz}\ \emph {et~al.}(2019)\citenamefont
  {Yankowitz}, \citenamefont {Chen}, \citenamefont {Polshyn}, \citenamefont
  {Zhang}, \citenamefont {Watanabe}, \citenamefont {Taniguchi}, \citenamefont
  {Graf}, \citenamefont {Young},\ and\ \citenamefont
  {Dean}}]{yankowitz2019tuning}%
  \BibitemOpen
  \bibfield  {author} {\bibinfo {author} {\bibfnamefont {M.}~\bibnamefont
  {Yankowitz}}, \bibinfo {author} {\bibfnamefont {S.}~\bibnamefont {Chen}},
  \bibinfo {author} {\bibfnamefont {H.}~\bibnamefont {Polshyn}}, \bibinfo
  {author} {\bibfnamefont {Y.}~\bibnamefont {Zhang}}, \bibinfo {author}
  {\bibfnamefont {K.}~\bibnamefont {Watanabe}}, \bibinfo {author}
  {\bibfnamefont {T.}~\bibnamefont {Taniguchi}}, \bibinfo {author}
  {\bibfnamefont {D.}~\bibnamefont {Graf}}, \bibinfo {author} {\bibfnamefont
  {A.~F.}\ \bibnamefont {Young}},\ and\ \bibinfo {author} {\bibfnamefont
  {C.~R.}\ \bibnamefont {Dean}},\ }\bibfield  {title} {\bibinfo {title} {Tuning
  superconductivity in twisted bilayer graphene},\ }\href@noop {} {\bibfield
  {journal} {\bibinfo  {journal} {Science}\ }\textbf {\bibinfo {volume}
  {363}},\ \bibinfo {pages} {1059} (\bibinfo {year} {2019})}\BibitemShut
  {NoStop}%
\bibitem [{\citenamefont {Bistritzer}\ and\ \citenamefont
  {MacDonald}(2011)}]{bistritzer2011moire}%
  \BibitemOpen
  \bibfield  {author} {\bibinfo {author} {\bibfnamefont {R.}~\bibnamefont
  {Bistritzer}}\ and\ \bibinfo {author} {\bibfnamefont {A.~H.}\ \bibnamefont
  {MacDonald}},\ }\bibfield  {title} {\bibinfo {title} {Moir{\'e} bands in
  twisted double-layer graphene},\ }\href@noop {} {\bibfield  {journal}
  {\bibinfo  {journal} {Proceedings of the National Academy of Sciences}\
  }\textbf {\bibinfo {volume} {108}},\ \bibinfo {pages} {12233} (\bibinfo
  {year} {2011})}\BibitemShut {NoStop}%
\bibitem [{\citenamefont {Dos~Santos}\ \emph {et~al.}(2007)\citenamefont
  {Dos~Santos}, \citenamefont {Peres},\ and\ \citenamefont
  {Neto}}]{dos2007graphene}%
  \BibitemOpen
  \bibfield  {author} {\bibinfo {author} {\bibfnamefont {J.~L.}\ \bibnamefont
  {Dos~Santos}}, \bibinfo {author} {\bibfnamefont {N.}~\bibnamefont {Peres}},\
  and\ \bibinfo {author} {\bibfnamefont {A.~C.}\ \bibnamefont {Neto}},\
  }\bibfield  {title} {\bibinfo {title} {Graphene bilayer with a twist:
  Electronic structure},\ }\href@noop {} {\bibfield  {journal} {\bibinfo
  {journal} {Physical review letters}\ }\textbf {\bibinfo {volume} {99}},\
  \bibinfo {pages} {256802} (\bibinfo {year} {2007})}\BibitemShut {NoStop}%
\bibitem [{\citenamefont {Sharpe}\ \emph {et~al.}(2019)\citenamefont {Sharpe},
  \citenamefont {Fox}, \citenamefont {Barnard}, \citenamefont {Finney},
  \citenamefont {Watanabe}, \citenamefont {Taniguchi}, \citenamefont
  {Kastner},\ and\ \citenamefont {Goldhaber-Gordon}}]{sharpe2019emergent}%
  \BibitemOpen
  \bibfield  {author} {\bibinfo {author} {\bibfnamefont {A.~L.}\ \bibnamefont
  {Sharpe}}, \bibinfo {author} {\bibfnamefont {E.~J.}\ \bibnamefont {Fox}},
  \bibinfo {author} {\bibfnamefont {A.~W.}\ \bibnamefont {Barnard}}, \bibinfo
  {author} {\bibfnamefont {J.}~\bibnamefont {Finney}}, \bibinfo {author}
  {\bibfnamefont {K.}~\bibnamefont {Watanabe}}, \bibinfo {author}
  {\bibfnamefont {T.}~\bibnamefont {Taniguchi}}, \bibinfo {author}
  {\bibfnamefont {M.}~\bibnamefont {Kastner}},\ and\ \bibinfo {author}
  {\bibfnamefont {D.}~\bibnamefont {Goldhaber-Gordon}},\ }\bibfield  {title}
  {\bibinfo {title} {Emergent ferromagnetism near three-quarters filling in
  twisted bilayer graphene},\ }\href@noop {} {\bibfield  {journal} {\bibinfo
  {journal} {Science}\ }\textbf {\bibinfo {volume} {365}},\ \bibinfo {pages}
  {605} (\bibinfo {year} {2019})}\BibitemShut {NoStop}%
\bibitem [{\citenamefont {Zhang}\ \emph
  {et~al.}(2019{\natexlab{a}})\citenamefont {Zhang}, \citenamefont {Mao},\ and\
  \citenamefont {Senthil}}]{zhang2019twisted}%
  \BibitemOpen
  \bibfield  {author} {\bibinfo {author} {\bibfnamefont {Y.-H.}\ \bibnamefont
  {Zhang}}, \bibinfo {author} {\bibfnamefont {D.}~\bibnamefont {Mao}},\ and\
  \bibinfo {author} {\bibfnamefont {T.}~\bibnamefont {Senthil}},\ }\bibfield
  {title} {\bibinfo {title} {Twisted bilayer graphene aligned with hexagonal
  boron nitride: anomalous hall effect and a lattice model},\ }\href@noop {}
  {\bibfield  {journal} {\bibinfo  {journal} {Physical Review Research}\
  }\textbf {\bibinfo {volume} {1}},\ \bibinfo {pages} {033126} (\bibinfo {year}
  {2019}{\natexlab{a}})}\BibitemShut {NoStop}%
\bibitem [{\citenamefont {Bultinck}\ \emph {et~al.}(2020)\citenamefont
  {Bultinck}, \citenamefont {Khalaf}, \citenamefont {Liu}, \citenamefont
  {Chatterjee}, \citenamefont {Vishwanath},\ and\ \citenamefont
  {Zaletel}}]{bultinck2020ground}%
  \BibitemOpen
  \bibfield  {author} {\bibinfo {author} {\bibfnamefont {N.}~\bibnamefont
  {Bultinck}}, \bibinfo {author} {\bibfnamefont {E.}~\bibnamefont {Khalaf}},
  \bibinfo {author} {\bibfnamefont {S.}~\bibnamefont {Liu}}, \bibinfo {author}
  {\bibfnamefont {S.}~\bibnamefont {Chatterjee}}, \bibinfo {author}
  {\bibfnamefont {A.}~\bibnamefont {Vishwanath}},\ and\ \bibinfo {author}
  {\bibfnamefont {M.~P.}\ \bibnamefont {Zaletel}},\ }\bibfield  {title}
  {\bibinfo {title} {Ground state and hidden symmetry of magic-angle graphene
  at even integer filling},\ }\href
  {https://doi.org/10.1103/PhysRevX.10.031034} {\bibfield  {journal} {\bibinfo
  {journal} {Phys. Rev. X}\ }\textbf {\bibinfo {volume} {10}},\ \bibinfo
  {pages} {031034} (\bibinfo {year} {2020})}\BibitemShut {NoStop}%
\bibitem [{\citenamefont {Song}\ \emph {et~al.}(2019)\citenamefont {Song},
  \citenamefont {Wang}, \citenamefont {Shi}, \citenamefont {Li}, \citenamefont
  {Fang},\ and\ \citenamefont {Bernevig}}]{song2019all}%
  \BibitemOpen
  \bibfield  {author} {\bibinfo {author} {\bibfnamefont {Z.}~\bibnamefont
  {Song}}, \bibinfo {author} {\bibfnamefont {Z.}~\bibnamefont {Wang}}, \bibinfo
  {author} {\bibfnamefont {W.}~\bibnamefont {Shi}}, \bibinfo {author}
  {\bibfnamefont {G.}~\bibnamefont {Li}}, \bibinfo {author} {\bibfnamefont
  {C.}~\bibnamefont {Fang}},\ and\ \bibinfo {author} {\bibfnamefont {B.~A.}\
  \bibnamefont {Bernevig}},\ }\bibfield  {title} {\bibinfo {title} {All magic
  angles in twisted bilayer graphene are topological},\ }\href@noop {}
  {\bibfield  {journal} {\bibinfo  {journal} {Physical review letters}\
  }\textbf {\bibinfo {volume} {123}},\ \bibinfo {pages} {036401} (\bibinfo
  {year} {2019})}\BibitemShut {NoStop}%
\bibitem [{\citenamefont {Song}\ \emph {et~al.}(2021)\citenamefont {Song},
  \citenamefont {Lian}, \citenamefont {Regnault},\ and\ \citenamefont
  {Bernevig}}]{song2021twisted}%
  \BibitemOpen
  \bibfield  {author} {\bibinfo {author} {\bibfnamefont {Z.-D.}\ \bibnamefont
  {Song}}, \bibinfo {author} {\bibfnamefont {B.}~\bibnamefont {Lian}}, \bibinfo
  {author} {\bibfnamefont {N.}~\bibnamefont {Regnault}},\ and\ \bibinfo
  {author} {\bibfnamefont {B.~A.}\ \bibnamefont {Bernevig}},\ }\bibfield
  {title} {\bibinfo {title} {Twisted bilayer graphene. ii. stable symmetry
  anomaly},\ }\href {https://doi.org/10.1103/PhysRevB.103.205412} {\bibfield
  {journal} {\bibinfo  {journal} {Phys. Rev. B}\ }\textbf {\bibinfo {volume}
  {103}},\ \bibinfo {pages} {205412} (\bibinfo {year} {2021})}\BibitemShut
  {NoStop}%
\bibitem [{\citenamefont {Lian}\ \emph {et~al.}(2021)\citenamefont {Lian},
  \citenamefont {Song}, \citenamefont {Regnault}, \citenamefont {Efetov},
  \citenamefont {Yazdani},\ and\ \citenamefont {Bernevig}}]{lian2021twisted}%
  \BibitemOpen
  \bibfield  {author} {\bibinfo {author} {\bibfnamefont {B.}~\bibnamefont
  {Lian}}, \bibinfo {author} {\bibfnamefont {Z.-D.}\ \bibnamefont {Song}},
  \bibinfo {author} {\bibfnamefont {N.}~\bibnamefont {Regnault}}, \bibinfo
  {author} {\bibfnamefont {D.~K.}\ \bibnamefont {Efetov}}, \bibinfo {author}
  {\bibfnamefont {A.}~\bibnamefont {Yazdani}},\ and\ \bibinfo {author}
  {\bibfnamefont {B.~A.}\ \bibnamefont {Bernevig}},\ }\bibfield  {title}
  {\bibinfo {title} {Twisted bilayer graphene. iv. exact insulator ground
  states and phase diagram},\ }\href
  {https://doi.org/10.1103/PhysRevB.103.205414} {\bibfield  {journal} {\bibinfo
   {journal} {Phys. Rev. B}\ }\textbf {\bibinfo {volume} {103}},\ \bibinfo
  {pages} {205414} (\bibinfo {year} {2021})}\BibitemShut {NoStop}%
\bibitem [{\citenamefont {Shen}\ \emph {et~al.}(2021)\citenamefont {Shen},
  \citenamefont {Ying}, \citenamefont {Liu}, \citenamefont {Liu}, \citenamefont
  {Li}, \citenamefont {Wang}, \citenamefont {Tang}, \citenamefont {Zhao},
  \citenamefont {Chu}, \citenamefont {Watanabe}, \citenamefont {Taniguchi},
  \citenamefont {Yang}, \citenamefont {Shi}, \citenamefont {Qu}, \citenamefont
  {Lu}, \citenamefont {Yang},\ and\ \citenamefont {Zhang}}]{shen2021emergence}%
  \BibitemOpen
  \bibfield  {author} {\bibinfo {author} {\bibfnamefont {C.}~\bibnamefont
  {Shen}}, \bibinfo {author} {\bibfnamefont {J.}~\bibnamefont {Ying}}, \bibinfo
  {author} {\bibfnamefont {L.}~\bibnamefont {Liu}}, \bibinfo {author}
  {\bibfnamefont {J.}~\bibnamefont {Liu}}, \bibinfo {author} {\bibfnamefont
  {N.}~\bibnamefont {Li}}, \bibinfo {author} {\bibfnamefont {S.}~\bibnamefont
  {Wang}}, \bibinfo {author} {\bibfnamefont {J.}~\bibnamefont {Tang}}, \bibinfo
  {author} {\bibfnamefont {Y.}~\bibnamefont {Zhao}}, \bibinfo {author}
  {\bibfnamefont {Y.}~\bibnamefont {Chu}}, \bibinfo {author} {\bibfnamefont
  {K.}~\bibnamefont {Watanabe}}, \bibinfo {author} {\bibfnamefont
  {T.}~\bibnamefont {Taniguchi}}, \bibinfo {author} {\bibfnamefont
  {R.}~\bibnamefont {Yang}}, \bibinfo {author} {\bibfnamefont {D.}~\bibnamefont
  {Shi}}, \bibinfo {author} {\bibfnamefont {F.}~\bibnamefont {Qu}}, \bibinfo
  {author} {\bibfnamefont {L.}~\bibnamefont {Lu}}, \bibinfo {author}
  {\bibfnamefont {W.}~\bibnamefont {Yang}},\ and\ \bibinfo {author}
  {\bibfnamefont {G.}~\bibnamefont {Zhang}},\ }\bibfield  {title} {\bibinfo
  {title} {Emergence of chern insulating states in non-magic angle twisted
  bilayer graphene},\ }\href {https://doi.org/10.1088/0256-307X/38/4/047301}
  {\bibfield  {journal} {\bibinfo  {journal} {Chinese Physics Letters}\
  }\textbf {\bibinfo {volume} {38}},\ \bibinfo {eid} {047301} (\bibinfo {year}
  {2021})}\BibitemShut {NoStop}%
\bibitem [{\citenamefont {Abouelkomsan}\ \emph {et~al.}(2020)\citenamefont
  {Abouelkomsan}, \citenamefont {Liu},\ and\ \citenamefont
  {Bergholtz}}]{abouelkomsan2020particle}%
  \BibitemOpen
  \bibfield  {author} {\bibinfo {author} {\bibfnamefont {A.}~\bibnamefont
  {Abouelkomsan}}, \bibinfo {author} {\bibfnamefont {Z.}~\bibnamefont {Liu}},\
  and\ \bibinfo {author} {\bibfnamefont {E.~J.}\ \bibnamefont {Bergholtz}},\
  }\bibfield  {title} {\bibinfo {title} {Particle-hole duality, emergent fermi
  liquids, and fractional chern insulators in moir{\'e} flatbands},\
  }\href@noop {} {\bibfield  {journal} {\bibinfo  {journal} {Physical review
  letters}\ }\textbf {\bibinfo {volume} {124}},\ \bibinfo {pages} {106803}
  (\bibinfo {year} {2020})}\BibitemShut {NoStop}%
\bibitem [{\citenamefont {Repellin}\ and\ \citenamefont
  {Senthil}(2020)}]{repellin2020chern}%
  \BibitemOpen
  \bibfield  {author} {\bibinfo {author} {\bibfnamefont {C.}~\bibnamefont
  {Repellin}}\ and\ \bibinfo {author} {\bibfnamefont {T.}~\bibnamefont
  {Senthil}},\ }\bibfield  {title} {\bibinfo {title} {Chern bands of twisted
  bilayer graphene: Fractional chern insulators and spin phase transition},\
  }\href@noop {} {\bibfield  {journal} {\bibinfo  {journal} {Physical Review
  Research}\ }\textbf {\bibinfo {volume} {2}},\ \bibinfo {pages} {023238}
  (\bibinfo {year} {2020})}\BibitemShut {NoStop}%
\bibitem [{\citenamefont {Wilhelm}\ \emph {et~al.}(2021)\citenamefont
  {Wilhelm}, \citenamefont {Lang},\ and\ \citenamefont
  {L{\"a}uchli}}]{wilhelm2021interplay}%
  \BibitemOpen
  \bibfield  {author} {\bibinfo {author} {\bibfnamefont {P.}~\bibnamefont
  {Wilhelm}}, \bibinfo {author} {\bibfnamefont {T.~C.}\ \bibnamefont {Lang}},\
  and\ \bibinfo {author} {\bibfnamefont {A.~M.}\ \bibnamefont {L{\"a}uchli}},\
  }\bibfield  {title} {\bibinfo {title} {Interplay of fractional chern
  insulator and charge density wave phases in twisted bilayer graphene},\
  }\href@noop {} {\bibfield  {journal} {\bibinfo  {journal} {Physical Review
  B}\ }\textbf {\bibinfo {volume} {103}},\ \bibinfo {pages} {125406} (\bibinfo
  {year} {2021})}\BibitemShut {NoStop}%
\bibitem [{\citenamefont {Xie}\ \emph {et~al.}(2021)\citenamefont {Xie},
  \citenamefont {Pierce}, \citenamefont {Park}, \citenamefont {Parker},
  \citenamefont {Khalaf}, \citenamefont {Ledwith}, \citenamefont {Cao},
  \citenamefont {Lee}, \citenamefont {Chen}, \citenamefont {Forrester} \emph
  {et~al.}}]{xie2021fractional}%
  \BibitemOpen
  \bibfield  {author} {\bibinfo {author} {\bibfnamefont {Y.}~\bibnamefont
  {Xie}}, \bibinfo {author} {\bibfnamefont {A.~T.}\ \bibnamefont {Pierce}},
  \bibinfo {author} {\bibfnamefont {J.~M.}\ \bibnamefont {Park}}, \bibinfo
  {author} {\bibfnamefont {D.~E.}\ \bibnamefont {Parker}}, \bibinfo {author}
  {\bibfnamefont {E.}~\bibnamefont {Khalaf}}, \bibinfo {author} {\bibfnamefont
  {P.}~\bibnamefont {Ledwith}}, \bibinfo {author} {\bibfnamefont
  {Y.}~\bibnamefont {Cao}}, \bibinfo {author} {\bibfnamefont {S.~H.}\
  \bibnamefont {Lee}}, \bibinfo {author} {\bibfnamefont {S.}~\bibnamefont
  {Chen}}, \bibinfo {author} {\bibfnamefont {P.~R.}\ \bibnamefont {Forrester}},
  \emph {et~al.},\ }\bibfield  {title} {\bibinfo {title} {Fractional chern
  insulators in magic-angle twisted bilayer graphene},\ }\href@noop {}
  {\bibfield  {journal} {\bibinfo  {journal} {Nature}\ }\textbf {\bibinfo
  {volume} {600}},\ \bibinfo {pages} {439} (\bibinfo {year}
  {2021})}\BibitemShut {NoStop}%
\bibitem [{\citenamefont {Mora}\ \emph {et~al.}(2019)\citenamefont {Mora},
  \citenamefont {Regnault},\ and\ \citenamefont
  {Bernevig}}]{mora2019flatbands}%
  \BibitemOpen
  \bibfield  {author} {\bibinfo {author} {\bibfnamefont {C.}~\bibnamefont
  {Mora}}, \bibinfo {author} {\bibfnamefont {N.}~\bibnamefont {Regnault}},\
  and\ \bibinfo {author} {\bibfnamefont {B.~A.}\ \bibnamefont {Bernevig}},\
  }\bibfield  {title} {\bibinfo {title} {Flatbands and perfect metal in
  trilayer moir{\'e} graphene},\ }\href@noop {} {\bibfield  {journal} {\bibinfo
   {journal} {Physical review letters}\ }\textbf {\bibinfo {volume} {123}},\
  \bibinfo {pages} {026402} (\bibinfo {year} {2019})}\BibitemShut {NoStop}%
\bibitem [{\citenamefont {Cea}\ \emph {et~al.}(2019)\citenamefont {Cea},
  \citenamefont {Walet},\ and\ \citenamefont {Guinea}}]{cea2019twists}%
  \BibitemOpen
  \bibfield  {author} {\bibinfo {author} {\bibfnamefont {T.}~\bibnamefont
  {Cea}}, \bibinfo {author} {\bibfnamefont {N.~R.}\ \bibnamefont {Walet}},\
  and\ \bibinfo {author} {\bibfnamefont {F.}~\bibnamefont {Guinea}},\
  }\bibfield  {title} {\bibinfo {title} {Twists and the electronic structure of
  graphitic materials},\ }\href@noop {} {\bibfield  {journal} {\bibinfo
  {journal} {Nano letters}\ }\textbf {\bibinfo {volume} {19}},\ \bibinfo
  {pages} {8683} (\bibinfo {year} {2019})}\BibitemShut {NoStop}%
\bibitem [{\citenamefont {Zhu}\ \emph {et~al.}(2020)\citenamefont {Zhu},
  \citenamefont {Carr}, \citenamefont {Massatt}, \citenamefont {Luskin},\ and\
  \citenamefont {Kaxiras}}]{zhu2020twisted}%
  \BibitemOpen
  \bibfield  {author} {\bibinfo {author} {\bibfnamefont {Z.}~\bibnamefont
  {Zhu}}, \bibinfo {author} {\bibfnamefont {S.}~\bibnamefont {Carr}}, \bibinfo
  {author} {\bibfnamefont {D.}~\bibnamefont {Massatt}}, \bibinfo {author}
  {\bibfnamefont {M.}~\bibnamefont {Luskin}},\ and\ \bibinfo {author}
  {\bibfnamefont {E.}~\bibnamefont {Kaxiras}},\ }\bibfield  {title} {\bibinfo
  {title} {Twisted trilayer graphene: A precisely tunable platform for
  correlated electrons},\ }\href@noop {} {\bibfield  {journal} {\bibinfo
  {journal} {Physical review letters}\ }\textbf {\bibinfo {volume} {125}},\
  \bibinfo {pages} {116404} (\bibinfo {year} {2020})}\BibitemShut {NoStop}%
\bibitem [{\citenamefont {Mao}\ \emph {et~al.}(2023)\citenamefont {Mao},
  \citenamefont {Guerci},\ and\ \citenamefont {Mora}}]{mao2023supermoire}%
  \BibitemOpen
  \bibfield  {author} {\bibinfo {author} {\bibfnamefont {Y.}~\bibnamefont
  {Mao}}, \bibinfo {author} {\bibfnamefont {D.}~\bibnamefont {Guerci}},\ and\
  \bibinfo {author} {\bibfnamefont {C.}~\bibnamefont {Mora}},\ }\bibfield
  {title} {\bibinfo {title} {Supermoir{\'e} low-energy effective theory of
  twisted trilayer graphene},\ }\href@noop {} {\bibfield  {journal} {\bibinfo
  {journal} {Physical Review B}\ }\textbf {\bibinfo {volume} {107}},\ \bibinfo
  {pages} {125423} (\bibinfo {year} {2023})}\BibitemShut {NoStop}%
\bibitem [{\citenamefont {Lin}\ \emph {et~al.}(2022)\citenamefont {Lin},
  \citenamefont {Li}, \citenamefont {Su},\ and\ \citenamefont
  {Ni}}]{lin2022energetic}%
  \BibitemOpen
  \bibfield  {author} {\bibinfo {author} {\bibfnamefont {X.}~\bibnamefont
  {Lin}}, \bibinfo {author} {\bibfnamefont {C.}~\bibnamefont {Li}}, \bibinfo
  {author} {\bibfnamefont {K.}~\bibnamefont {Su}},\ and\ \bibinfo {author}
  {\bibfnamefont {J.}~\bibnamefont {Ni}},\ }\bibfield  {title} {\bibinfo
  {title} {Energetic stability and spatial inhomogeneity in the local
  electronic structure of relaxed twisted trilayer graphene},\ }\href@noop {}
  {\bibfield  {journal} {\bibinfo  {journal} {Physical Review B}\ }\textbf
  {\bibinfo {volume} {106}},\ \bibinfo {pages} {075423} (\bibinfo {year}
  {2022})}\BibitemShut {NoStop}%
\bibitem [{\citenamefont {Ma}\ \emph {et~al.}(2023)\citenamefont {Ma},
  \citenamefont {Li}, \citenamefont {Lu}, \citenamefont {Xu}, \citenamefont
  {Gao},\ and\ \citenamefont {Xie}}]{ma2023doubled}%
  \BibitemOpen
  \bibfield  {author} {\bibinfo {author} {\bibfnamefont {Z.}~\bibnamefont
  {Ma}}, \bibinfo {author} {\bibfnamefont {S.}~\bibnamefont {Li}}, \bibinfo
  {author} {\bibfnamefont {M.}~\bibnamefont {Lu}}, \bibinfo {author}
  {\bibfnamefont {D.-H.}\ \bibnamefont {Xu}}, \bibinfo {author} {\bibfnamefont
  {J.-H.}\ \bibnamefont {Gao}},\ and\ \bibinfo {author} {\bibfnamefont
  {X.}~\bibnamefont {Xie}},\ }\bibfield  {title} {\bibinfo {title} {Doubled
  moir{\'e} flat bands in double-twisted few-layer graphite},\ }\href@noop {}
  {\bibfield  {journal} {\bibinfo  {journal} {Science China Physics, Mechanics
  \& Astronomy}\ }\textbf {\bibinfo {volume} {66}},\ \bibinfo {pages} {227211}
  (\bibinfo {year} {2023})}\BibitemShut {NoStop}%
\bibitem [{\citenamefont {Hao}\ \emph {et~al.}(2021)\citenamefont {Hao},
  \citenamefont {Zimmerman}, \citenamefont {Ledwith}, \citenamefont {Khalaf},
  \citenamefont {Najafabadi}, \citenamefont {Watanabe}, \citenamefont
  {Taniguchi}, \citenamefont {Vishwanath},\ and\ \citenamefont
  {Kim}}]{hao2021electric}%
  \BibitemOpen
  \bibfield  {author} {\bibinfo {author} {\bibfnamefont {Z.}~\bibnamefont
  {Hao}}, \bibinfo {author} {\bibfnamefont {A.}~\bibnamefont {Zimmerman}},
  \bibinfo {author} {\bibfnamefont {P.}~\bibnamefont {Ledwith}}, \bibinfo
  {author} {\bibfnamefont {E.}~\bibnamefont {Khalaf}}, \bibinfo {author}
  {\bibfnamefont {D.~H.}\ \bibnamefont {Najafabadi}}, \bibinfo {author}
  {\bibfnamefont {K.}~\bibnamefont {Watanabe}}, \bibinfo {author}
  {\bibfnamefont {T.}~\bibnamefont {Taniguchi}}, \bibinfo {author}
  {\bibfnamefont {A.}~\bibnamefont {Vishwanath}},\ and\ \bibinfo {author}
  {\bibfnamefont {P.}~\bibnamefont {Kim}},\ }\bibfield  {title} {\bibinfo
  {title} {Electric field--tunable superconductivity in alternating-twist
  magic-angle trilayer graphene},\ }\href@noop {} {\bibfield  {journal}
  {\bibinfo  {journal} {Science}\ }\textbf {\bibinfo {volume} {371}},\ \bibinfo
  {pages} {1133} (\bibinfo {year} {2021})}\BibitemShut {NoStop}%
\bibitem [{\citenamefont {Lake}\ and\ \citenamefont
  {Senthil}(2021)}]{lake2021reentrant}%
  \BibitemOpen
  \bibfield  {author} {\bibinfo {author} {\bibfnamefont {E.}~\bibnamefont
  {Lake}}\ and\ \bibinfo {author} {\bibfnamefont {T.}~\bibnamefont {Senthil}},\
  }\bibfield  {title} {\bibinfo {title} {Reentrant superconductivity through a
  quantum lifshitz transition in twisted trilayer graphene},\ }\href
  {https://doi.org/10.1103/PhysRevB.104.174505} {\bibfield  {journal} {\bibinfo
   {journal} {Phys. Rev. B}\ }\textbf {\bibinfo {volume} {104}},\ \bibinfo
  {pages} {174505} (\bibinfo {year} {2021})}\BibitemShut {NoStop}%
\bibitem [{\citenamefont {Qin}\ and\ \citenamefont
  {MacDonald}(2021)}]{qin2021plane}%
  \BibitemOpen
  \bibfield  {author} {\bibinfo {author} {\bibfnamefont {W.}~\bibnamefont
  {Qin}}\ and\ \bibinfo {author} {\bibfnamefont {A.~H.}\ \bibnamefont
  {MacDonald}},\ }\bibfield  {title} {\bibinfo {title} {In-plane critical
  magnetic fields in magic-angle twisted trilayer graphene},\ }\href@noop {}
  {\bibfield  {journal} {\bibinfo  {journal} {Physical Review Letters}\
  }\textbf {\bibinfo {volume} {127}},\ \bibinfo {pages} {097001} (\bibinfo
  {year} {2021})}\BibitemShut {NoStop}%
\bibitem [{\citenamefont {Park}\ \emph {et~al.}(2022)\citenamefont {Park},
  \citenamefont {Cao}, \citenamefont {Xia}, \citenamefont {Sun}, \citenamefont
  {Watanabe}, \citenamefont {Taniguchi},\ and\ \citenamefont
  {Jarillo-Herrero}}]{park2022robust}%
  \BibitemOpen
  \bibfield  {author} {\bibinfo {author} {\bibfnamefont {J.~M.}\ \bibnamefont
  {Park}}, \bibinfo {author} {\bibfnamefont {Y.}~\bibnamefont {Cao}}, \bibinfo
  {author} {\bibfnamefont {L.-Q.}\ \bibnamefont {Xia}}, \bibinfo {author}
  {\bibfnamefont {S.}~\bibnamefont {Sun}}, \bibinfo {author} {\bibfnamefont
  {K.}~\bibnamefont {Watanabe}}, \bibinfo {author} {\bibfnamefont
  {T.}~\bibnamefont {Taniguchi}},\ and\ \bibinfo {author} {\bibfnamefont
  {P.}~\bibnamefont {Jarillo-Herrero}},\ }\bibfield  {title} {\bibinfo {title}
  {Robust superconductivity in magic-angle multilayer graphene family},\
  }\href@noop {} {\bibfield  {journal} {\bibinfo  {journal} {Nature Materials}\
  }\textbf {\bibinfo {volume} {21}},\ \bibinfo {pages} {877} (\bibinfo {year}
  {2022})}\BibitemShut {NoStop}%
\bibitem [{\citenamefont {Zhang}\ \emph {et~al.}(2022)\citenamefont {Zhang},
  \citenamefont {Polski}, \citenamefont {Lewandowski}, \citenamefont {Thomson},
  \citenamefont {Peng}, \citenamefont {Choi}, \citenamefont {Kim},
  \citenamefont {Watanabe}, \citenamefont {Taniguchi}, \citenamefont {Alicea}
  \emph {et~al.}}]{zhang2022promotion}%
  \BibitemOpen
  \bibfield  {author} {\bibinfo {author} {\bibfnamefont {Y.}~\bibnamefont
  {Zhang}}, \bibinfo {author} {\bibfnamefont {R.}~\bibnamefont {Polski}},
  \bibinfo {author} {\bibfnamefont {C.}~\bibnamefont {Lewandowski}}, \bibinfo
  {author} {\bibfnamefont {A.}~\bibnamefont {Thomson}}, \bibinfo {author}
  {\bibfnamefont {Y.}~\bibnamefont {Peng}}, \bibinfo {author} {\bibfnamefont
  {Y.}~\bibnamefont {Choi}}, \bibinfo {author} {\bibfnamefont {H.}~\bibnamefont
  {Kim}}, \bibinfo {author} {\bibfnamefont {K.}~\bibnamefont {Watanabe}},
  \bibinfo {author} {\bibfnamefont {T.}~\bibnamefont {Taniguchi}}, \bibinfo
  {author} {\bibfnamefont {J.}~\bibnamefont {Alicea}}, \emph {et~al.},\
  }\bibfield  {title} {\bibinfo {title} {Promotion of superconductivity in
  magic-angle graphene multilayers},\ }\href@noop {} {\bibfield  {journal}
  {\bibinfo  {journal} {Science}\ }\textbf {\bibinfo {volume} {377}},\ \bibinfo
  {pages} {1538} (\bibinfo {year} {2022})}\BibitemShut {NoStop}%
\bibitem [{\citenamefont {Ledwith}\ \emph {et~al.}(2021)\citenamefont
  {Ledwith}, \citenamefont {Khalaf}, \citenamefont {Zhu}, \citenamefont {Carr},
  \citenamefont {Kaxiras},\ and\ \citenamefont {Vishwanath}}]{ledwith2021tb}%
  \BibitemOpen
  \bibfield  {author} {\bibinfo {author} {\bibfnamefont {P.~J.}\ \bibnamefont
  {Ledwith}}, \bibinfo {author} {\bibfnamefont {E.}~\bibnamefont {Khalaf}},
  \bibinfo {author} {\bibfnamefont {Z.}~\bibnamefont {Zhu}}, \bibinfo {author}
  {\bibfnamefont {S.}~\bibnamefont {Carr}}, \bibinfo {author} {\bibfnamefont
  {E.}~\bibnamefont {Kaxiras}},\ and\ \bibinfo {author} {\bibfnamefont
  {A.}~\bibnamefont {Vishwanath}},\ }\bibfield  {title} {\bibinfo {title} {Tb
  or not tb? contrasting properties of twisted bilayer graphene and the
  alternating twist $ n $-layer structures ($ n= 3, 4, 5,\backslash\dots $)},\
  }\href@noop {} {\bibfield  {journal} {\bibinfo  {journal} {arXiv preprint
  arXiv:2111.11060}\ } (\bibinfo {year} {2021})}\BibitemShut {NoStop}%
\bibitem [{\citenamefont {Khalaf}\ \emph {et~al.}(2019)\citenamefont {Khalaf},
  \citenamefont {Kruchkov}, \citenamefont {Tarnopolsky},\ and\ \citenamefont
  {Vishwanath}}]{khalaf2019magic}%
  \BibitemOpen
  \bibfield  {author} {\bibinfo {author} {\bibfnamefont {E.}~\bibnamefont
  {Khalaf}}, \bibinfo {author} {\bibfnamefont {A.~J.}\ \bibnamefont
  {Kruchkov}}, \bibinfo {author} {\bibfnamefont {G.}~\bibnamefont
  {Tarnopolsky}},\ and\ \bibinfo {author} {\bibfnamefont {A.}~\bibnamefont
  {Vishwanath}},\ }\bibfield  {title} {\bibinfo {title} {Magic angle hierarchy
  in twisted graphene multilayers},\ }\href@noop {} {\bibfield  {journal}
  {\bibinfo  {journal} {Physical Review B}\ }\textbf {\bibinfo {volume}
  {100}},\ \bibinfo {pages} {085109} (\bibinfo {year} {2019})}\BibitemShut
  {NoStop}%
\bibitem [{\citenamefont {Popov}\ and\ \citenamefont
  {Tarnopolsky}(2023)}]{popov2023magic}%
  \BibitemOpen
  \bibfield  {author} {\bibinfo {author} {\bibfnamefont {F.~K.}\ \bibnamefont
  {Popov}}\ and\ \bibinfo {author} {\bibfnamefont {G.}~\bibnamefont
  {Tarnopolsky}},\ }\bibfield  {title} {\bibinfo {title} {Magic angles in
  equal-twist trilayer graphene},\ }\href@noop {} {\bibfield  {journal}
  {\bibinfo  {journal} {arXiv preprint arXiv:2303.15505}\ } (\bibinfo {year}
  {2023})}\BibitemShut {NoStop}%
\bibitem [{\citenamefont {Chen}\ \emph {et~al.}(2019)\citenamefont {Chen},
  \citenamefont {Sharpe}, \citenamefont {Gallagher}, \citenamefont {Rosen},
  \citenamefont {Fox}, \citenamefont {Jiang}, \citenamefont {Lyu},
  \citenamefont {Li}, \citenamefont {Watanabe}, \citenamefont {Taniguchi} \emph
  {et~al.}}]{chen2019signatures}%
  \BibitemOpen
  \bibfield  {author} {\bibinfo {author} {\bibfnamefont {G.}~\bibnamefont
  {Chen}}, \bibinfo {author} {\bibfnamefont {A.~L.}\ \bibnamefont {Sharpe}},
  \bibinfo {author} {\bibfnamefont {P.}~\bibnamefont {Gallagher}}, \bibinfo
  {author} {\bibfnamefont {I.~T.}\ \bibnamefont {Rosen}}, \bibinfo {author}
  {\bibfnamefont {E.~J.}\ \bibnamefont {Fox}}, \bibinfo {author} {\bibfnamefont
  {L.}~\bibnamefont {Jiang}}, \bibinfo {author} {\bibfnamefont
  {B.}~\bibnamefont {Lyu}}, \bibinfo {author} {\bibfnamefont {H.}~\bibnamefont
  {Li}}, \bibinfo {author} {\bibfnamefont {K.}~\bibnamefont {Watanabe}},
  \bibinfo {author} {\bibfnamefont {T.}~\bibnamefont {Taniguchi}}, \emph
  {et~al.},\ }\bibfield  {title} {\bibinfo {title} {Signatures of tunable
  superconductivity in a trilayer graphene moir{\'e} superlattice},\
  }\href@noop {} {\bibfield  {journal} {\bibinfo  {journal} {Nature}\ }\textbf
  {\bibinfo {volume} {572}},\ \bibinfo {pages} {215} (\bibinfo {year}
  {2019})}\BibitemShut {NoStop}%
\bibitem [{\citenamefont {Chen}\ \emph {et~al.}(2020)\citenamefont {Chen},
  \citenamefont {Sharpe}, \citenamefont {Fox}, \citenamefont {Zhang},
  \citenamefont {Wang}, \citenamefont {Jiang}, \citenamefont {Lyu},
  \citenamefont {Li}, \citenamefont {Watanabe}, \citenamefont {Taniguchi} \emph
  {et~al.}}]{chen2020tunable}%
  \BibitemOpen
  \bibfield  {author} {\bibinfo {author} {\bibfnamefont {G.}~\bibnamefont
  {Chen}}, \bibinfo {author} {\bibfnamefont {A.~L.}\ \bibnamefont {Sharpe}},
  \bibinfo {author} {\bibfnamefont {E.~J.}\ \bibnamefont {Fox}}, \bibinfo
  {author} {\bibfnamefont {Y.-H.}\ \bibnamefont {Zhang}}, \bibinfo {author}
  {\bibfnamefont {S.}~\bibnamefont {Wang}}, \bibinfo {author} {\bibfnamefont
  {L.}~\bibnamefont {Jiang}}, \bibinfo {author} {\bibfnamefont
  {B.}~\bibnamefont {Lyu}}, \bibinfo {author} {\bibfnamefont {H.}~\bibnamefont
  {Li}}, \bibinfo {author} {\bibfnamefont {K.}~\bibnamefont {Watanabe}},
  \bibinfo {author} {\bibfnamefont {T.}~\bibnamefont {Taniguchi}}, \emph
  {et~al.},\ }\bibfield  {title} {\bibinfo {title} {Tunable correlated chern
  insulator and ferromagnetism in a moir{\'e} superlattice},\ }\href@noop {}
  {\bibfield  {journal} {\bibinfo  {journal} {Nature}\ }\textbf {\bibinfo
  {volume} {579}},\ \bibinfo {pages} {56} (\bibinfo {year} {2020})}\BibitemShut
  {NoStop}%
\bibitem [{\citenamefont {Zhang}\ \emph
  {et~al.}(2019{\natexlab{b}})\citenamefont {Zhang}, \citenamefont {Mao},
  \citenamefont {Cao}, \citenamefont {Jarillo-Herrero},\ and\ \citenamefont
  {Senthil}}]{zhang2019nearly}%
  \BibitemOpen
  \bibfield  {author} {\bibinfo {author} {\bibfnamefont {Y.-H.}\ \bibnamefont
  {Zhang}}, \bibinfo {author} {\bibfnamefont {D.}~\bibnamefont {Mao}}, \bibinfo
  {author} {\bibfnamefont {Y.}~\bibnamefont {Cao}}, \bibinfo {author}
  {\bibfnamefont {P.}~\bibnamefont {Jarillo-Herrero}},\ and\ \bibinfo {author}
  {\bibfnamefont {T.}~\bibnamefont {Senthil}},\ }\bibfield  {title} {\bibinfo
  {title} {Nearly flat chern bands in moir{\'e} superlattices},\ }\href@noop {}
  {\bibfield  {journal} {\bibinfo  {journal} {Physical Review B}\ }\textbf
  {\bibinfo {volume} {99}},\ \bibinfo {pages} {075127} (\bibinfo {year}
  {2019}{\natexlab{b}})}\BibitemShut {NoStop}%
\bibitem [{\citenamefont {Liu}\ \emph {et~al.}(2022)\citenamefont {Liu},
  \citenamefont {Shi}, \citenamefont {Yang},\ and\ \citenamefont
  {Zhang}}]{liu2022magic}%
  \BibitemOpen
  \bibfield  {author} {\bibinfo {author} {\bibfnamefont {Z.}~\bibnamefont
  {Liu}}, \bibinfo {author} {\bibfnamefont {W.}~\bibnamefont {Shi}}, \bibinfo
  {author} {\bibfnamefont {T.}~\bibnamefont {Yang}},\ and\ \bibinfo {author}
  {\bibfnamefont {Z.}~\bibnamefont {Zhang}},\ }\bibfield  {title} {\bibinfo
  {title} {Magic angles and flat chern bands in alternating-twist multilayer
  graphene system},\ }\href@noop {} {\bibfield  {journal} {\bibinfo  {journal}
  {Journal of Materials Science \& Technology}\ }\textbf {\bibinfo {volume}
  {111}},\ \bibinfo {pages} {28} (\bibinfo {year} {2022})}\BibitemShut
  {NoStop}%
\bibitem [{\citenamefont {Xie}\ \emph {et~al.}(2022)\citenamefont {Xie},
  \citenamefont {Peng}, \citenamefont {Zhang},\ and\ \citenamefont
  {Liu}}]{xie2022alternating}%
  \BibitemOpen
  \bibfield  {author} {\bibinfo {author} {\bibfnamefont {B.}~\bibnamefont
  {Xie}}, \bibinfo {author} {\bibfnamefont {R.}~\bibnamefont {Peng}}, \bibinfo
  {author} {\bibfnamefont {S.}~\bibnamefont {Zhang}},\ and\ \bibinfo {author}
  {\bibfnamefont {J.}~\bibnamefont {Liu}},\ }\bibfield  {title} {\bibinfo
  {title} {Alternating twisted multilayer graphene: generic partition rules,
  double flat bands, and orbital magnetoelectric effect},\ }\href@noop {}
  {\bibfield  {journal} {\bibinfo  {journal} {npj Computational Materials}\
  }\textbf {\bibinfo {volume} {8}},\ \bibinfo {pages} {1} (\bibinfo {year}
  {2022})}\BibitemShut {NoStop}%
\bibitem [{\citenamefont {Liu}\ \emph {et~al.}(2020)\citenamefont {Liu},
  \citenamefont {Hao}, \citenamefont {Khalaf}, \citenamefont {Lee},
  \citenamefont {Ronen}, \citenamefont {Yoo}, \citenamefont {Haei~Najafabadi},
  \citenamefont {Watanabe}, \citenamefont {Taniguchi}, \citenamefont
  {Vishwanath} \emph {et~al.}}]{liu2020tunable}%
  \BibitemOpen
  \bibfield  {author} {\bibinfo {author} {\bibfnamefont {X.}~\bibnamefont
  {Liu}}, \bibinfo {author} {\bibfnamefont {Z.}~\bibnamefont {Hao}}, \bibinfo
  {author} {\bibfnamefont {E.}~\bibnamefont {Khalaf}}, \bibinfo {author}
  {\bibfnamefont {J.~Y.}\ \bibnamefont {Lee}}, \bibinfo {author} {\bibfnamefont
  {Y.}~\bibnamefont {Ronen}}, \bibinfo {author} {\bibfnamefont
  {H.}~\bibnamefont {Yoo}}, \bibinfo {author} {\bibfnamefont {D.}~\bibnamefont
  {Haei~Najafabadi}}, \bibinfo {author} {\bibfnamefont {K.}~\bibnamefont
  {Watanabe}}, \bibinfo {author} {\bibfnamefont {T.}~\bibnamefont {Taniguchi}},
  \bibinfo {author} {\bibfnamefont {A.}~\bibnamefont {Vishwanath}}, \emph
  {et~al.},\ }\bibfield  {title} {\bibinfo {title} {Tunable spin-polarized
  correlated states in twisted double bilayer graphene},\ }\href@noop {}
  {\bibfield  {journal} {\bibinfo  {journal} {Nature}\ }\textbf {\bibinfo
  {volume} {583}},\ \bibinfo {pages} {221} (\bibinfo {year}
  {2020})}\BibitemShut {NoStop}%
\bibitem [{\citenamefont {Lee}\ \emph {et~al.}(2019)\citenamefont {Lee},
  \citenamefont {Khalaf}, \citenamefont {Liu}, \citenamefont {Liu},
  \citenamefont {Hao}, \citenamefont {Kim},\ and\ \citenamefont
  {Vishwanath}}]{lee2019theory}%
  \BibitemOpen
  \bibfield  {author} {\bibinfo {author} {\bibfnamefont {J.~Y.}\ \bibnamefont
  {Lee}}, \bibinfo {author} {\bibfnamefont {E.}~\bibnamefont {Khalaf}},
  \bibinfo {author} {\bibfnamefont {S.}~\bibnamefont {Liu}}, \bibinfo {author}
  {\bibfnamefont {X.}~\bibnamefont {Liu}}, \bibinfo {author} {\bibfnamefont
  {Z.}~\bibnamefont {Hao}}, \bibinfo {author} {\bibfnamefont {P.}~\bibnamefont
  {Kim}},\ and\ \bibinfo {author} {\bibfnamefont {A.}~\bibnamefont
  {Vishwanath}},\ }\bibfield  {title} {\bibinfo {title} {Theory of correlated
  insulating behaviour and spin-triplet superconductivity in twisted double
  bilayer graphene},\ }\href@noop {} {\bibfield  {journal} {\bibinfo  {journal}
  {Nature communications}\ }\textbf {\bibinfo {volume} {10}},\ \bibinfo {pages}
  {5333} (\bibinfo {year} {2019})}\BibitemShut {NoStop}%
\bibitem [{\citenamefont {Cao}\ \emph {et~al.}(2020)\citenamefont {Cao},
  \citenamefont {Rodan-Legrain}, \citenamefont {Rubies-Bigorda}, \citenamefont
  {Park}, \citenamefont {Watanabe}, \citenamefont {Taniguchi},\ and\
  \citenamefont {Jarillo-Herrero}}]{cao2020tunable}%
  \BibitemOpen
  \bibfield  {author} {\bibinfo {author} {\bibfnamefont {Y.}~\bibnamefont
  {Cao}}, \bibinfo {author} {\bibfnamefont {D.}~\bibnamefont {Rodan-Legrain}},
  \bibinfo {author} {\bibfnamefont {O.}~\bibnamefont {Rubies-Bigorda}},
  \bibinfo {author} {\bibfnamefont {J.~M.}\ \bibnamefont {Park}}, \bibinfo
  {author} {\bibfnamefont {K.}~\bibnamefont {Watanabe}}, \bibinfo {author}
  {\bibfnamefont {T.}~\bibnamefont {Taniguchi}},\ and\ \bibinfo {author}
  {\bibfnamefont {P.}~\bibnamefont {Jarillo-Herrero}},\ }\bibfield  {title}
  {\bibinfo {title} {Tunable correlated states and spin-polarized phases in
  twisted bilayer--bilayer graphene},\ }\href@noop {} {\bibfield  {journal}
  {\bibinfo  {journal} {Nature}\ }\textbf {\bibinfo {volume} {583}},\ \bibinfo
  {pages} {215} (\bibinfo {year} {2020})}\BibitemShut {NoStop}%
\bibitem [{\citenamefont {Shen}\ \emph {et~al.}(2020)\citenamefont {Shen},
  \citenamefont {Chu}, \citenamefont {Wu}, \citenamefont {Li}, \citenamefont
  {Wang}, \citenamefont {Zhao}, \citenamefont {Tang}, \citenamefont {Liu},
  \citenamefont {Tian}, \citenamefont {Watanabe} \emph
  {et~al.}}]{shen2020correlated}%
  \BibitemOpen
  \bibfield  {author} {\bibinfo {author} {\bibfnamefont {C.}~\bibnamefont
  {Shen}}, \bibinfo {author} {\bibfnamefont {Y.}~\bibnamefont {Chu}}, \bibinfo
  {author} {\bibfnamefont {Q.}~\bibnamefont {Wu}}, \bibinfo {author}
  {\bibfnamefont {N.}~\bibnamefont {Li}}, \bibinfo {author} {\bibfnamefont
  {S.}~\bibnamefont {Wang}}, \bibinfo {author} {\bibfnamefont {Y.}~\bibnamefont
  {Zhao}}, \bibinfo {author} {\bibfnamefont {J.}~\bibnamefont {Tang}}, \bibinfo
  {author} {\bibfnamefont {J.}~\bibnamefont {Liu}}, \bibinfo {author}
  {\bibfnamefont {J.}~\bibnamefont {Tian}}, \bibinfo {author} {\bibfnamefont
  {K.}~\bibnamefont {Watanabe}}, \emph {et~al.},\ }\bibfield  {title} {\bibinfo
  {title} {Correlated states in twisted double bilayer graphene},\ }\href@noop
  {} {\bibfield  {journal} {\bibinfo  {journal} {Nature Physics}\ }\textbf
  {\bibinfo {volume} {16}},\ \bibinfo {pages} {520} (\bibinfo {year}
  {2020})}\BibitemShut {NoStop}%
\bibitem [{\citenamefont {Oka}\ and\ \citenamefont {Aoki}(2009)}]{Oka_2009}%
  \BibitemOpen
  \bibfield  {author} {\bibinfo {author} {\bibfnamefont {T.}~\bibnamefont
  {Oka}}\ and\ \bibinfo {author} {\bibfnamefont {H.}~\bibnamefont {Aoki}},\
  }\bibfield  {title} {\bibinfo {title} {Photovoltaic hall effect in
  graphene},\ }\href {https://doi.org/10.1103/PhysRevB.79.081406} {\bibfield
  {journal} {\bibinfo  {journal} {Phys. Rev. B}\ }\textbf {\bibinfo {volume}
  {79}},\ \bibinfo {pages} {081406(R)} (\bibinfo {year} {2009})}\BibitemShut
  {NoStop}%
\bibitem [{\citenamefont {Gu}\ \emph {et~al.}(2011)\citenamefont {Gu},
  \citenamefont {Fertig}, \citenamefont {Arovas},\ and\ \citenamefont
  {Auerbach}}]{Gu_2011}%
  \BibitemOpen
  \bibfield  {author} {\bibinfo {author} {\bibfnamefont {Z.}~\bibnamefont
  {Gu}}, \bibinfo {author} {\bibfnamefont {H.~A.}\ \bibnamefont {Fertig}},
  \bibinfo {author} {\bibfnamefont {D.~P.}\ \bibnamefont {Arovas}},\ and\
  \bibinfo {author} {\bibfnamefont {A.}~\bibnamefont {Auerbach}},\ }\bibfield
  {title} {\bibinfo {title} {Floquet spectrum and transport through an
  irradiated graphene ribbon},\ }\href
  {https://doi.org/10.1103/PhysRevLett.107.216601} {\bibfield  {journal}
  {\bibinfo  {journal} {Phys. Rev. Lett.}\ }\textbf {\bibinfo {volume} {107}},\
  \bibinfo {pages} {216601} (\bibinfo {year} {2011})}\BibitemShut {NoStop}%
\bibitem [{\citenamefont {Kitagawa}\ \emph {et~al.}(2011)\citenamefont
  {Kitagawa}, \citenamefont {Oka}, \citenamefont {Brataas}, \citenamefont
  {Fu},\ and\ \citenamefont {Demler}}]{Kitagawa_2011}%
  \BibitemOpen
  \bibfield  {author} {\bibinfo {author} {\bibfnamefont {T.}~\bibnamefont
  {Kitagawa}}, \bibinfo {author} {\bibfnamefont {T.}~\bibnamefont {Oka}},
  \bibinfo {author} {\bibfnamefont {A.}~\bibnamefont {Brataas}}, \bibinfo
  {author} {\bibfnamefont {L.}~\bibnamefont {Fu}},\ and\ \bibinfo {author}
  {\bibfnamefont {E.}~\bibnamefont {Demler}},\ }\bibfield  {title} {\bibinfo
  {title} {Transport properties of nonequilibrium systems under the application
  of light: Photoinduced quantum hall insulators without landau levels},\
  }\href {https://doi.org/10.1103/PhysRevB.84.235108} {\bibfield  {journal}
  {\bibinfo  {journal} {Phys. Rev. B}\ }\textbf {\bibinfo {volume} {84}},\
  \bibinfo {pages} {235108} (\bibinfo {year} {2011})}\BibitemShut {NoStop}%
\bibitem [{\citenamefont {D\'ora}\ \emph {et~al.}(2012)\citenamefont {D\'ora},
  \citenamefont {Cayssol}, \citenamefont {Simon},\ and\ \citenamefont
  {Moessner}}]{Dora_2012}%
  \BibitemOpen
  \bibfield  {author} {\bibinfo {author} {\bibfnamefont {B.}~\bibnamefont
  {D\'ora}}, \bibinfo {author} {\bibfnamefont {J.}~\bibnamefont {Cayssol}},
  \bibinfo {author} {\bibfnamefont {F.}~\bibnamefont {Simon}},\ and\ \bibinfo
  {author} {\bibfnamefont {R.}~\bibnamefont {Moessner}},\ }\bibfield  {title}
  {\bibinfo {title} {Optically engineering the topological properties of a spin
  hall insulator},\ }\href {https://doi.org/10.1103/PhysRevLett.108.056602}
  {\bibfield  {journal} {\bibinfo  {journal} {Phys. Rev. Lett.}\ }\textbf
  {\bibinfo {volume} {108}},\ \bibinfo {pages} {056602} (\bibinfo {year}
  {2012})}\BibitemShut {NoStop}%
\bibitem [{\citenamefont {Iadecola}\ \emph {et~al.}(2013)\citenamefont
  {Iadecola}, \citenamefont {Campbell}, \citenamefont {Chamon}, \citenamefont
  {Hou}, \citenamefont {Jackiw}, \citenamefont {Pi},\ and\ \citenamefont
  {Kusminskiy}}]{Iadecola_2013}%
  \BibitemOpen
  \bibfield  {author} {\bibinfo {author} {\bibfnamefont {T.}~\bibnamefont
  {Iadecola}}, \bibinfo {author} {\bibfnamefont {D.}~\bibnamefont {Campbell}},
  \bibinfo {author} {\bibfnamefont {C.}~\bibnamefont {Chamon}}, \bibinfo
  {author} {\bibfnamefont {C.-Y.}\ \bibnamefont {Hou}}, \bibinfo {author}
  {\bibfnamefont {R.}~\bibnamefont {Jackiw}}, \bibinfo {author} {\bibfnamefont
  {S.-Y.}\ \bibnamefont {Pi}},\ and\ \bibinfo {author} {\bibfnamefont {S.~V.}\
  \bibnamefont {Kusminskiy}},\ }\bibfield  {title} {\bibinfo {title} {Materials
  design from nonequilibrium steady states: Driven graphene as a tunable
  semiconductor with topological properties},\ }\href
  {https://doi.org/10.1103/PhysRevLett.110.176603} {\bibfield  {journal}
  {\bibinfo  {journal} {Phys. Rev. Lett.}\ }\textbf {\bibinfo {volume} {110}},\
  \bibinfo {pages} {176603} (\bibinfo {year} {2013})}\BibitemShut {NoStop}%
\bibitem [{\citenamefont {Syzranov}\ \emph {et~al.}(2013)\citenamefont
  {Syzranov}, \citenamefont {Rodionov}, \citenamefont {Kugel},\ and\
  \citenamefont {Nori}}]{Syzranov_2013}%
  \BibitemOpen
  \bibfield  {author} {\bibinfo {author} {\bibfnamefont {S.~V.}\ \bibnamefont
  {Syzranov}}, \bibinfo {author} {\bibfnamefont {Y.~I.}\ \bibnamefont
  {Rodionov}}, \bibinfo {author} {\bibfnamefont {K.~I.}\ \bibnamefont
  {Kugel}},\ and\ \bibinfo {author} {\bibfnamefont {F.}~\bibnamefont {Nori}},\
  }\bibfield  {title} {\bibinfo {title} {Strongly anisotropic dirac
  quasiparticles in irradiated graphene},\ }\href
  {https://doi.org/10.1103/PhysRevB.88.241112} {\bibfield  {journal} {\bibinfo
  {journal} {Phys. Rev. B}\ }\textbf {\bibinfo {volume} {88}},\ \bibinfo
  {pages} {241112(R)} (\bibinfo {year} {2013})}\BibitemShut {NoStop}%
\bibitem [{\citenamefont {Ezawa}(2013)}]{Ezawa_2013}%
  \BibitemOpen
  \bibfield  {author} {\bibinfo {author} {\bibfnamefont {M.}~\bibnamefont
  {Ezawa}},\ }\bibfield  {title} {\bibinfo {title} {Photoinduced topological
  phase transition and a single dirac-cone state in silicene},\ }\href
  {https://doi.org/10.1103/PhysRevLett.110.026603} {\bibfield  {journal}
  {\bibinfo  {journal} {Phys. Rev. Lett.}\ }\textbf {\bibinfo {volume} {110}},\
  \bibinfo {pages} {026603} (\bibinfo {year} {2013})}\BibitemShut {NoStop}%
\bibitem [{\citenamefont {Kundu}\ \emph {et~al.}(2014)\citenamefont {Kundu},
  \citenamefont {Fertig},\ and\ \citenamefont {Seradjeh}}]{Kundu_2014}%
  \BibitemOpen
  \bibfield  {author} {\bibinfo {author} {\bibfnamefont {A.}~\bibnamefont
  {Kundu}}, \bibinfo {author} {\bibfnamefont {H.}~\bibnamefont {Fertig}},\ and\
  \bibinfo {author} {\bibfnamefont {B.}~\bibnamefont {Seradjeh}},\ }\bibfield
  {title} {\bibinfo {title} {Effective theory of floquet topological
  transitions},\ }\href@noop {} {\bibfield  {journal} {\bibinfo  {journal}
  {Physical review letters}\ }\textbf {\bibinfo {volume} {113}},\ \bibinfo
  {pages} {236803} (\bibinfo {year} {2014})}\BibitemShut {NoStop}%
\bibitem [{\citenamefont {Grushin}\ \emph {et~al.}(2014)\citenamefont
  {Grushin}, \citenamefont {G{\'o}mez-Le{\'o}n},\ and\ \citenamefont
  {Neupert}}]{Grushin_2014}%
  \BibitemOpen
  \bibfield  {author} {\bibinfo {author} {\bibfnamefont {A.~G.}\ \bibnamefont
  {Grushin}}, \bibinfo {author} {\bibfnamefont {{\'A}.}~\bibnamefont
  {G{\'o}mez-Le{\'o}n}},\ and\ \bibinfo {author} {\bibfnamefont
  {T.}~\bibnamefont {Neupert}},\ }\bibfield  {title} {\bibinfo {title} {Floquet
  fractional chern insulators},\ }\href@noop {} {\bibfield  {journal} {\bibinfo
   {journal} {Physical review letters}\ }\textbf {\bibinfo {volume} {112}},\
  \bibinfo {pages} {156801} (\bibinfo {year} {2014})}\BibitemShut {NoStop}%
\bibitem [{\citenamefont {Kundu}\ \emph {et~al.}(2016)\citenamefont {Kundu},
  \citenamefont {Fertig},\ and\ \citenamefont {Seradjeh}}]{Kundu_2016}%
  \BibitemOpen
  \bibfield  {author} {\bibinfo {author} {\bibfnamefont {A.}~\bibnamefont
  {Kundu}}, \bibinfo {author} {\bibfnamefont {H.~A.}\ \bibnamefont {Fertig}},\
  and\ \bibinfo {author} {\bibfnamefont {B.}~\bibnamefont {Seradjeh}},\
  }\bibfield  {title} {\bibinfo {title} {Floquet-engineered valleytronics in
  dirac systems},\ }\href {https://doi.org/10.1103/PhysRevLett.116.016802}
  {\bibfield  {journal} {\bibinfo  {journal} {Phys. Rev. Lett.}\ }\textbf
  {\bibinfo {volume} {116}},\ \bibinfo {pages} {016802} (\bibinfo {year}
  {2016})}\BibitemShut {NoStop}%
\bibitem [{\citenamefont {Rudner}\ and\ \citenamefont
  {Lindner}(2020)}]{rudner2020band}%
  \BibitemOpen
  \bibfield  {author} {\bibinfo {author} {\bibfnamefont {M.~S.}\ \bibnamefont
  {Rudner}}\ and\ \bibinfo {author} {\bibfnamefont {N.~H.}\ \bibnamefont
  {Lindner}},\ }\bibfield  {title} {\bibinfo {title} {Band structure
  engineering and non-equilibrium dynamics in floquet topological insulators},\
  }\href@noop {} {\bibfield  {journal} {\bibinfo  {journal} {Nature reviews
  physics}\ }\textbf {\bibinfo {volume} {2}},\ \bibinfo {pages} {229} (\bibinfo
  {year} {2020})}\BibitemShut {NoStop}%
\bibitem [{\citenamefont {Sentef}\ \emph {et~al.}(2015)\citenamefont {Sentef},
  \citenamefont {Claassen}, \citenamefont {Kemper}, \citenamefont {Moritz},
  \citenamefont {Oka}, \citenamefont {Freericks},\ and\ \citenamefont
  {Devereaux}}]{sentef2015theory}%
  \BibitemOpen
  \bibfield  {author} {\bibinfo {author} {\bibfnamefont {M.}~\bibnamefont
  {Sentef}}, \bibinfo {author} {\bibfnamefont {M.}~\bibnamefont {Claassen}},
  \bibinfo {author} {\bibfnamefont {A.}~\bibnamefont {Kemper}}, \bibinfo
  {author} {\bibfnamefont {B.}~\bibnamefont {Moritz}}, \bibinfo {author}
  {\bibfnamefont {T.}~\bibnamefont {Oka}}, \bibinfo {author} {\bibfnamefont
  {J.}~\bibnamefont {Freericks}},\ and\ \bibinfo {author} {\bibfnamefont
  {T.}~\bibnamefont {Devereaux}},\ }\bibfield  {title} {\bibinfo {title}
  {Theory of floquet band formation and local pseudospin textures in pump-probe
  photoemission of graphene},\ }\href@noop {} {\bibfield  {journal} {\bibinfo
  {journal} {Nature communications}\ }\textbf {\bibinfo {volume} {6}},\
  \bibinfo {pages} {7047} (\bibinfo {year} {2015})}\BibitemShut {NoStop}%
\bibitem [{\citenamefont {Li}\ \emph {et~al.}(2020)\citenamefont {Li},
  \citenamefont {Fertig},\ and\ \citenamefont {Seradjeh}}]{li2020floquet}%
  \BibitemOpen
  \bibfield  {author} {\bibinfo {author} {\bibfnamefont {Y.}~\bibnamefont
  {Li}}, \bibinfo {author} {\bibfnamefont {H.}~\bibnamefont {Fertig}},\ and\
  \bibinfo {author} {\bibfnamefont {B.}~\bibnamefont {Seradjeh}},\ }\bibfield
  {title} {\bibinfo {title} {Floquet-engineered topological flat bands in
  irradiated twisted bilayer graphene},\ }\href@noop {} {\bibfield  {journal}
  {\bibinfo  {journal} {Physical Review Research}\ }\textbf {\bibinfo {volume}
  {2}},\ \bibinfo {pages} {043275} (\bibinfo {year} {2020})}\BibitemShut
  {NoStop}%
\bibitem [{\citenamefont {Yang}\ \emph {et~al.}(2023)\citenamefont {Yang},
  \citenamefont {Esin}, \citenamefont {Lewandowski},\ and\ \citenamefont
  {Refael}}]{yang2023optical}%
  \BibitemOpen
  \bibfield  {author} {\bibinfo {author} {\bibfnamefont {C.}~\bibnamefont
  {Yang}}, \bibinfo {author} {\bibfnamefont {I.}~\bibnamefont {Esin}}, \bibinfo
  {author} {\bibfnamefont {C.}~\bibnamefont {Lewandowski}},\ and\ \bibinfo
  {author} {\bibfnamefont {G.}~\bibnamefont {Refael}},\ }\bibfield  {title}
  {\bibinfo {title} {Optical control of slow topological electrons in moir\'e
  systems},\ }\href {https://doi.org/10.1103/PhysRevLett.131.026901} {\bibfield
   {journal} {\bibinfo  {journal} {Phys. Rev. Lett.}\ }\textbf {\bibinfo
  {volume} {131}},\ \bibinfo {pages} {026901} (\bibinfo {year}
  {2023})}\BibitemShut {NoStop}%
\bibitem [{\citenamefont {Hu}\ \emph {et~al.}(2023)\citenamefont {Hu},
  \citenamefont {Zhou},\ and\ \citenamefont {Liu}}]{hu2022floquet}%
  \BibitemOpen
  \bibfield  {author} {\bibinfo {author} {\bibfnamefont {P.-S.}\ \bibnamefont
  {Hu}}, \bibinfo {author} {\bibfnamefont {Y.-H.}\ \bibnamefont {Zhou}},\ and\
  \bibinfo {author} {\bibfnamefont {Z.}~\bibnamefont {Liu}},\ }\bibfield
  {title} {\bibinfo {title} {Floquet fractional chern insulators and competing
  phases in twisted bilayer graphene},\ }\href@noop {} {\bibfield  {journal}
  {\bibinfo  {journal} {SciPost Physics}\ }\textbf {\bibinfo {volume} {15}},\
  \bibinfo {pages} {148} (\bibinfo {year} {2023})}\BibitemShut {NoStop}%
\bibitem [{\citenamefont {Vogl}\ \emph
  {et~al.}(2020{\natexlab{a}})\citenamefont {Vogl}, \citenamefont
  {Rodriguez-Vega},\ and\ \citenamefont {Fiete}}]{vogl2020effective}%
  \BibitemOpen
  \bibfield  {author} {\bibinfo {author} {\bibfnamefont {M.}~\bibnamefont
  {Vogl}}, \bibinfo {author} {\bibfnamefont {M.}~\bibnamefont
  {Rodriguez-Vega}},\ and\ \bibinfo {author} {\bibfnamefont {G.~A.}\
  \bibnamefont {Fiete}},\ }\bibfield  {title} {\bibinfo {title} {Effective
  floquet hamiltonians for periodically driven twisted bilayer graphene},\
  }\href@noop {} {\bibfield  {journal} {\bibinfo  {journal} {Physical Review
  B}\ }\textbf {\bibinfo {volume} {101}},\ \bibinfo {pages} {235411} (\bibinfo
  {year} {2020}{\natexlab{a}})}\BibitemShut {NoStop}%
\bibitem [{\citenamefont {Vogl}\ \emph
  {et~al.}(2020{\natexlab{b}})\citenamefont {Vogl}, \citenamefont
  {Rodriguez-Vega},\ and\ \citenamefont {Fiete}}]{vogl2020floquet}%
  \BibitemOpen
  \bibfield  {author} {\bibinfo {author} {\bibfnamefont {M.}~\bibnamefont
  {Vogl}}, \bibinfo {author} {\bibfnamefont {M.}~\bibnamefont
  {Rodriguez-Vega}},\ and\ \bibinfo {author} {\bibfnamefont {G.~A.}\
  \bibnamefont {Fiete}},\ }\bibfield  {title} {\bibinfo {title} {Floquet
  engineering of interlayer couplings: Tuning the magic angle of twisted
  bilayer graphene at the exit of a waveguide},\ }\href@noop {} {\bibfield
  {journal} {\bibinfo  {journal} {Physical Review B}\ }\textbf {\bibinfo
  {volume} {101}},\ \bibinfo {pages} {241408(R)} (\bibinfo {year}
  {2020}{\natexlab{b}})}\BibitemShut {NoStop}%
\bibitem [{\citenamefont {Katz}\ \emph {et~al.}(2020)\citenamefont {Katz},
  \citenamefont {Refael},\ and\ \citenamefont {Lindner}}]{katz2020optically}%
  \BibitemOpen
  \bibfield  {author} {\bibinfo {author} {\bibfnamefont {O.}~\bibnamefont
  {Katz}}, \bibinfo {author} {\bibfnamefont {G.}~\bibnamefont {Refael}},\ and\
  \bibinfo {author} {\bibfnamefont {N.~H.}\ \bibnamefont {Lindner}},\
  }\bibfield  {title} {\bibinfo {title} {Optically induced flat bands in
  twisted bilayer graphene},\ }\href@noop {} {\bibfield  {journal} {\bibinfo
  {journal} {Physical Review B}\ }\textbf {\bibinfo {volume} {102}},\ \bibinfo
  {pages} {155123} (\bibinfo {year} {2020})}\BibitemShut {NoStop}%
\bibitem [{\citenamefont {Topp}\ \emph {et~al.}(2019)\citenamefont {Topp},
  \citenamefont {Jotzu}, \citenamefont {McIver}, \citenamefont {Xian},
  \citenamefont {Rubio},\ and\ \citenamefont {Sentef}}]{topp2019topological}%
  \BibitemOpen
  \bibfield  {author} {\bibinfo {author} {\bibfnamefont {G.~E.}\ \bibnamefont
  {Topp}}, \bibinfo {author} {\bibfnamefont {G.}~\bibnamefont {Jotzu}},
  \bibinfo {author} {\bibfnamefont {J.~W.}\ \bibnamefont {McIver}}, \bibinfo
  {author} {\bibfnamefont {L.}~\bibnamefont {Xian}}, \bibinfo {author}
  {\bibfnamefont {A.}~\bibnamefont {Rubio}},\ and\ \bibinfo {author}
  {\bibfnamefont {M.~A.}\ \bibnamefont {Sentef}},\ }\bibfield  {title}
  {\bibinfo {title} {Topological floquet engineering of twisted bilayer
  graphene},\ }\href {https://doi.org/10.1103/PhysRevResearch.1.023031}
  {\bibfield  {journal} {\bibinfo  {journal} {Phys. Rev. Res.}\ }\textbf
  {\bibinfo {volume} {1}},\ \bibinfo {pages} {023031} (\bibinfo {year}
  {2019})}\BibitemShut {NoStop}%
\bibitem [{\citenamefont {Topp}\ \emph {et~al.}(2021)\citenamefont {Topp},
  \citenamefont {Eckhardt}, \citenamefont {Kennes}, \citenamefont {Sentef},\
  and\ \citenamefont {T\"orm\"a}}]{topp2021light}%
  \BibitemOpen
  \bibfield  {author} {\bibinfo {author} {\bibfnamefont {G.~E.}\ \bibnamefont
  {Topp}}, \bibinfo {author} {\bibfnamefont {C.~J.}\ \bibnamefont {Eckhardt}},
  \bibinfo {author} {\bibfnamefont {D.~M.}\ \bibnamefont {Kennes}}, \bibinfo
  {author} {\bibfnamefont {M.~A.}\ \bibnamefont {Sentef}},\ and\ \bibinfo
  {author} {\bibfnamefont {P.}~\bibnamefont {T\"orm\"a}},\ }\bibfield  {title}
  {\bibinfo {title} {Light-matter coupling and quantum geometry in moir\'e
  materials},\ }\href {https://doi.org/10.1103/PhysRevB.104.064306} {\bibfield
  {journal} {\bibinfo  {journal} {Phys. Rev. B}\ }\textbf {\bibinfo {volume}
  {104}},\ \bibinfo {pages} {064306} (\bibinfo {year} {2021})}\BibitemShut
  {NoStop}%
\bibitem [{\citenamefont {Rodriguez-Vega}\ \emph {et~al.}(2021)\citenamefont
  {Rodriguez-Vega}, \citenamefont {Vogl},\ and\ \citenamefont
  {Fiete}}]{rodriguez2021low}%
  \BibitemOpen
  \bibfield  {author} {\bibinfo {author} {\bibfnamefont {M.}~\bibnamefont
  {Rodriguez-Vega}}, \bibinfo {author} {\bibfnamefont {M.}~\bibnamefont
  {Vogl}},\ and\ \bibinfo {author} {\bibfnamefont {G.~A.}\ \bibnamefont
  {Fiete}},\ }\bibfield  {title} {\bibinfo {title} {Low-frequency and
  moir{\'e}--floquet engineering: A review},\ }\href@noop {} {\bibfield
  {journal} {\bibinfo  {journal} {Annals of Physics}\ }\textbf {\bibinfo
  {volume} {435}},\ \bibinfo {pages} {168434} (\bibinfo {year}
  {2021})}\BibitemShut {NoStop}%
\bibitem [{\citenamefont {Rodriguez-Vega}\ \emph {et~al.}(2020)\citenamefont
  {Rodriguez-Vega}, \citenamefont {Vogl},\ and\ \citenamefont
  {Fiete}}]{rodriguez2020floquet}%
  \BibitemOpen
  \bibfield  {author} {\bibinfo {author} {\bibfnamefont {M.}~\bibnamefont
  {Rodriguez-Vega}}, \bibinfo {author} {\bibfnamefont {M.}~\bibnamefont
  {Vogl}},\ and\ \bibinfo {author} {\bibfnamefont {G.~A.}\ \bibnamefont
  {Fiete}},\ }\bibfield  {title} {\bibinfo {title} {Floquet engineering of
  twisted double bilayer graphene},\ }\href@noop {} {\bibfield  {journal}
  {\bibinfo  {journal} {Physical Review Research}\ }\textbf {\bibinfo {volume}
  {2}},\ \bibinfo {pages} {033494} (\bibinfo {year} {2020})}\BibitemShut
  {NoStop}%
\bibitem [{\citenamefont {Assi}\ \emph {et~al.}(2021)\citenamefont {Assi},
  \citenamefont {LeBlanc}, \citenamefont {Rodriguez-Vega}, \citenamefont
  {Bahlouli},\ and\ \citenamefont {Vogl}}]{assi2021floquet}%
  \BibitemOpen
  \bibfield  {author} {\bibinfo {author} {\bibfnamefont {I.}~\bibnamefont
  {Assi}}, \bibinfo {author} {\bibfnamefont {J.}~\bibnamefont {LeBlanc}},
  \bibinfo {author} {\bibfnamefont {M.}~\bibnamefont {Rodriguez-Vega}},
  \bibinfo {author} {\bibfnamefont {H.}~\bibnamefont {Bahlouli}},\ and\
  \bibinfo {author} {\bibfnamefont {M.}~\bibnamefont {Vogl}},\ }\bibfield
  {title} {\bibinfo {title} {Floquet engineering and nonequilibrium topological
  maps in twisted trilayer graphene},\ }\href@noop {} {\bibfield  {journal}
  {\bibinfo  {journal} {Physical Review B}\ }\textbf {\bibinfo {volume}
  {104}},\ \bibinfo {pages} {195429} (\bibinfo {year} {2021})}\BibitemShut
  {NoStop}%
\bibitem [{\citenamefont {Lu}\ \emph {et~al.}(2021)\citenamefont {Lu},
  \citenamefont {Zeng}, \citenamefont {Liu}, \citenamefont {Gao},\ and\
  \citenamefont {Xie}}]{lu2021valley}%
  \BibitemOpen
  \bibfield  {author} {\bibinfo {author} {\bibfnamefont {M.}~\bibnamefont
  {Lu}}, \bibinfo {author} {\bibfnamefont {J.}~\bibnamefont {Zeng}}, \bibinfo
  {author} {\bibfnamefont {H.}~\bibnamefont {Liu}}, \bibinfo {author}
  {\bibfnamefont {J.-H.}\ \bibnamefont {Gao}},\ and\ \bibinfo {author}
  {\bibfnamefont {X.}~\bibnamefont {Xie}},\ }\bibfield  {title} {\bibinfo
  {title} {Valley-selective floquet chern flat bands in twisted multilayer
  graphene},\ }\href@noop {} {\bibfield  {journal} {\bibinfo  {journal}
  {Physical Review B}\ }\textbf {\bibinfo {volume} {103}},\ \bibinfo {pages}
  {195146} (\bibinfo {year} {2021})}\BibitemShut {NoStop}%
\bibitem [{\citenamefont {Ledwith}\ \emph {et~al.}(2020)\citenamefont
  {Ledwith}, \citenamefont {Tarnopolsky}, \citenamefont {Khalaf},\ and\
  \citenamefont {Vishwanath}}]{ledwith2020fractional}%
  \BibitemOpen
  \bibfield  {author} {\bibinfo {author} {\bibfnamefont {P.~J.}\ \bibnamefont
  {Ledwith}}, \bibinfo {author} {\bibfnamefont {G.}~\bibnamefont
  {Tarnopolsky}}, \bibinfo {author} {\bibfnamefont {E.}~\bibnamefont
  {Khalaf}},\ and\ \bibinfo {author} {\bibfnamefont {A.}~\bibnamefont
  {Vishwanath}},\ }\bibfield  {title} {\bibinfo {title} {Fractional chern
  insulator states in twisted bilayer graphene: An analytical approach},\
  }\href {https://doi.org/10.1103/PhysRevResearch.2.023237} {\bibfield
  {journal} {\bibinfo  {journal} {Phys. Rev. Res.}\ }\textbf {\bibinfo {volume}
  {2}},\ \bibinfo {pages} {023237} (\bibinfo {year} {2020})}\BibitemShut
  {NoStop}%
\bibitem [{\citenamefont {Kuzmenko}\ \emph {et~al.}(2009)\citenamefont
  {Kuzmenko}, \citenamefont {Crassee}, \citenamefont {Van Der~Marel},
  \citenamefont {Blake},\ and\ \citenamefont
  {Novoselov}}]{kuzmenko2009determination}%
  \BibitemOpen
  \bibfield  {author} {\bibinfo {author} {\bibfnamefont {A.}~\bibnamefont
  {Kuzmenko}}, \bibinfo {author} {\bibfnamefont {I.}~\bibnamefont {Crassee}},
  \bibinfo {author} {\bibfnamefont {D.}~\bibnamefont {Van Der~Marel}}, \bibinfo
  {author} {\bibfnamefont {P.}~\bibnamefont {Blake}},\ and\ \bibinfo {author}
  {\bibfnamefont {K.}~\bibnamefont {Novoselov}},\ }\bibfield  {title} {\bibinfo
  {title} {Determination of the gate-tunable band gap and tight-binding
  parameters in bilayer graphene using infrared spectroscopy},\ }\href@noop {}
  {\bibfield  {journal} {\bibinfo  {journal} {Physical Review B}\ }\textbf
  {\bibinfo {volume} {80}},\ \bibinfo {pages} {165406} (\bibinfo {year}
  {2009})}\BibitemShut {NoStop}%
\bibitem [{\citenamefont {Leconte}\ \emph {et~al.}(2022)\citenamefont
  {Leconte}, \citenamefont {Park}, \citenamefont {An}, \citenamefont
  {Samudrala},\ and\ \citenamefont {Jung}}]{leconte2022electronic}%
  \BibitemOpen
  \bibfield  {author} {\bibinfo {author} {\bibfnamefont {N.}~\bibnamefont
  {Leconte}}, \bibinfo {author} {\bibfnamefont {Y.}~\bibnamefont {Park}},
  \bibinfo {author} {\bibfnamefont {J.}~\bibnamefont {An}}, \bibinfo {author}
  {\bibfnamefont {A.}~\bibnamefont {Samudrala}},\ and\ \bibinfo {author}
  {\bibfnamefont {J.}~\bibnamefont {Jung}},\ }\bibfield  {title} {\bibinfo
  {title} {Electronic structure of lattice relaxed alternating twist
  tng-multilayer graphene: from few layers to bulk at-graphite},\ }\href@noop
  {} {\bibfield  {journal} {\bibinfo  {journal} {2D Materials}\ }\textbf
  {\bibinfo {volume} {9}},\ \bibinfo {pages} {044002} (\bibinfo {year}
  {2022})}\BibitemShut {NoStop}%
\end{thebibliography}%

\end{document}